\documentclass[11pt,onecolumn]{IEEEtran}
\usepackage{subfigure}
\usepackage{epsfig,graphicx,psfrag}
\usepackage{amsfonts,amsmath,amssymb}
\newtheorem{lemma}{Lemma}
\newtheorem{theorem}{Theorem}
\newtheorem{cor}{Corollary}

\newtheorem{definition}{Definition}

\newcommand{\hGe}{\hat{G}^\epsilon}
\newcommand{\hEe}{\hat{E}^\epsilon}
\newcommand{\lf}{\left}
\newcommand{\rf}{\right}
\newtheorem{corollary}{Corollary}

\newcommand{\bone}{{\mathbf 1}}

\let\bl\bigl
\let\bbl\Bigl

\let\br\bigr
\let\bbr\Bigr

\newcommand{\I}{\mathcal{I}}
\newcommand{\E}{\mathbb{E}}
\newcommand{\R}{\mathbb{R}}

\newcommand{\cC}{\mathcal{C}}
\newcommand{\cA}{\mathcal{A}}

\newcommand{\N}{\mathbb{N}}
\newcommand{\Z}{\mathbb{Z}}

\newcommand{\cS}{\mathcal{S}}

\newcommand{\eg}{\emph{e.g.}}
\begin{document}
\title{\huge Product Multicommodity Flow in Wireless Networks}
\author{Ritesh Madan, Devavrat Shah, and Olivier Leveque
\thanks{R. Madan is at QUALCOMM-Flarion Technologies, D. Shah is with Departments of EECS and ESD at MIT, and Olivier
Leveque is at EPFL. Emails:~{ \tt rkmadan@stanfordalumni.org, devavrat@mit.edu,
olivier.leveque@epfl.ch.}
 }}
\maketitle


\begin{abstract}

We provide a tight approximate characterization of the
$n$-dimensional product multicommodity flow (PMF)
region for a wireless network of $n$ nodes. Separate
characterizations in terms of the spectral properties of
appropriate network graphs are obtained in both an information theoretic
sense and for a combinatorial interference model (\eg, Protocol
model). These provide an inner approximation to the $n^2$ dimensional
capacity region. These results answer the following
questions which arise naturally from previous work: (a) What is the
significance of $1/\sqrt{n}$ in the scaling laws for the Protocol
interference model obtained by Gupta and Kumar (2000)? (b) Can we
obtain a tight approximation to the ``maximum supportable flow" for
node distributions more general than the geometric random
distribution, traffic models other than randomly chosen
source-destination pairs, and under very general assumptions on the
channel fading model?

We first establish that the random source-destination model is
essentially a one-dimensional approximation to the capacity region,
and a special case of product multi-commodity flow. For
a wireline network (graph), a series of results starting from the result
of Leighton and Rao (1988) relate the product multicommodity flow to
the spectral (or cut) property of the graph. Building on these results,
for a combinatorial interference model given by a network
    and a conflict
graph, we relate the product multicommodity flow to the spectral
properties of the underlying graphs resulting in computational upper
and lower bounds. These results show that the $1/\sqrt{n}$ scaling
law obtained by Gupta and Kumar for a geometric random network can
be explained in terms of the combinatorial properties of a
geometric random network and the scaling law of the \emph{conductance}
of a grid graph. For the more interesting random fading
model with additive white Gaussian noise (AWGN), we show that the
scaling laws for PMF can again be tightly characterized by the
spectral properties of appropriately defined graphs. As an
implication, we obtain computationally efficient upper and lower
bounds on the PMF for any wireless network with a guaranteed
approximation factor.

\end{abstract}


%
\section{Introduction}

\subsection{Prior Work}\label{s:intro} \vspace{.1in}

An important open question in network information theory is that of
characterizing the capacity region of a wireless network of $n$
nodes, i.e., the set of all achievable rates between the $n^2$ pairs of
nodes in terms of the joint statistics of the channels between
these nodes. This has proved to be a very challenging question; even
the capacity of a relay network comprising of three nodes is
not known in complete generality.

Instead of trying to characterize the
capacity region for a general wireless network, the seminal paper by
Gupta and Kumar~\cite{GK} concentrated on obtaining the maximum achievable rate
for a particular communication model, geometric random distribution
of nodes, and randomly chosen source-destination pairs. They showed
that the maximum rate for the protocol interference model scales as
$\Theta(1/\sqrt{n})$ for $n$ nodes randomly placed on a sphere of
unit area. This precise characterization has been followed by many
interesting results for both combinatorial interference models and
the random fading information theoretic model for large random
networks; these include~\cite{GT, KV, EMPS_journal,F, GV, XXK} for
communication theoretic models, and~\cite{LT,XK} for
information theoretic results. These results are crucially based on
the assumption that a large number of nodes are randomly distributed
in a certain region, and the inherent symmetry in the random
source-destination pair traffic model.

Since the relative locations of wireless nodes play an
important role in the characterization of the capacity region, the
notion of transport capacity was defined in~\cite{GK}. A scaling law
for the transport capacity for the protocol interference model was
obtained in~\cite{GK}, while that for an information theoretic
setting was obtained in~\cite{XXK}. The transport capacity can be used
to obtain an upper bound on the achievable rate-region for
certain rate-tuples, but is not of much use in determining the feasibility of
a certain rate-tuple.
More
recently, information theoretic  outer bounds to the capacity
region of a wireless network with a finite number of nodes were
obtained in~\cite{AJV} for any wireless network, using the cut-set
bound~\cite[Ch. 14]{Cover_book}. We note that any achievable scheme
can be used to obtain a set of lower bounds. While the above is only
a discussion of a representative set of results in this area
(see~\cite{XK_mono} for a more detailed summary), we note that there
is no result which provides upper and lower bounds with a guaranteed
approximation factor for a general wireless network with a generic
random fading model. In this paper, we take the first steps towards
providing such a tight characterization under very general
assumptions. In doing so, we make  connections between
spectral graph theoretic results and network information theory.
This results in efficient methods to compute the above tight upper
and lower bounds.


\subsection{Contribution and Organization}
\vspace{.1in}

In Section~\ref{sec1}, we consider the product multicommodity flow (PMF)
as an $n$ dimensional approximation of the $n^2$ dimensional
capacity region. We show that the random souce-destination pair
traffic model is a special case of PMF and it is essentially a
one-dimensional approximation of the capacity region.

In Section~\ref{sec2}, we study the PMF for an arbitrary topology and a general
combinatorial interference model, of which the protocol model is a
special case. We show that the normalized cut capacity
(equivalently conductance) of a capacitated network graph induced by
the node placement and the interference model characterizes the PMF
(within $\log n$ factor). For this model, we also obtain a precise
scaling law for average delay using very elementary and almost
structure independent arguments.
We provide, possibly simpler, re-derivation of the (weaker by $\log^{2.5}n$ factor)
lower bound on the maximum flow obtained by Gupta and Kumar for randomly chosen
permutation flow on a geometric random graph with a protocol interference model.
Our derivation illustrates the connections between the combinatorial properties of geometric random
graphs and the maximum PMF.
While we do not discuss in detail in this
paper, the spectral properties of appropriate induced capacitated
graphs characterize the scaling laws obtained for mobile networks
in~\cite{GT}, \cite{DGT}.

In Section~\ref{sec3}, we address the question of characterizing the
PMF for a wireless network with Gaussian channels and random fading.
This is substantially more challenging than for the combinatorial
interference model because there is no obvious underlying network
graph that specifies the links which should be used for data
transmission. We construct a capacitated graph whose cut capacity
characterizes (in terms of tight upper and lower bounds) the PMF in
the wireless network. This construction allows one to use classical
network flow arguments to characterize and compute the PMF. We
illustrate the generality of our results by obtaining scaling laws for a
geometric random network and for a network where the number of
commodities is constant.

\section{Traffic Flows}\label{sec1}

In this section, we describe a class of traffic flows, namely,
the class of product multicommodity flows, that we study in this paper, and its
relevance. Consider a wireless network of $n$ nodes and denote the
node set as $V=\{1,\hdots,n\}$. A traffic matrix $\lambda =
[\lambda_{ij}] \in \R_+^{n\times n}$ is said to be {\em feasible},
if for each pair of nodes $(i,j), ~1\leq i, j \leq n$, data can be
transmitted from node $i$ to node $j$ at rate $\lambda_{ij}$. Note
that whether a traffic matrix $\lambda$ is feasible or not depends
on the model for the underlying wireless network, and we shall
describe the precise models for wireless networks in the later
sections.

We denote the capacity region by  $\Lambda$, i.e., $\Lambda$ is the
set of all feasible traffic matrices. Ideally, we would like to
characterize $\Lambda$. However, this is a hard problem in most cases. Instead, we
 characterize an {\em approximation} of $\Lambda$ under
general assumptions on the wireless network. For this, we consider
product multicommodity flow (PMF), defined as follows.

\vspace{.1in}

\begin{definition}[Product Multicommodity Flow (PMF)]
Let node $i$ be assigned a weight $\pi(i)$, for $1\leq i\leq n$.
Then the PMF corresponding to the weights $\pi\in \R_{+}^n$ and a flow rate $f\in \R_{+}$ is given by the function~\cite{LR99}
$M: \R^{n+1} \mapsto \R^{n \times n} $:
\[ M(f,\pi) = f \left[ \begin{array}{llll}
0 & \pi(1)\pi(2) &\hdots & \pi(1)\pi(n)\\
\pi(2)\pi(1) & 0 & \hdots & \pi(2)\pi(n)\\
\vdots & \vdots &\vdots &\vdots\\
\pi(n) \pi(1) & \pi(n)\pi(2)&\hdots& \ \ 0
\end{array}\right].
\]
\end{definition}
\vspace{.1in} The PMF is an $n$-dimensional approximation of the
$n^2$ dimensional capacity region $\Lambda$ with {\em product}
constraints. An important special case arises when
all the weights are 1, i.e., $\pi(i) = 1$ for $i=1,\hdots,n$. We call
such a flow \emph{uniform multicommodity flow (UMF)}. \vspace{.1in}

\begin{definition}[Uniform Multicommodity Flow (UMF)]
UMF with flow rate $f\in \R_{+}$ is given by $U(f) = f {\bf 1}$, where
${\bf 1}\in \mathbb{R}_+^{n\times n}$ is a matrix with all
entries equal to 1.
\end{definition}
\vspace{0.1in}
We denote by $f_\pi^*$ as the supremum over the flow rates for which the PMF
corresponding to the weights $\pi$ is feasible, i.e.,
\[ f^*_\pi = \sup \{ f\in R_{+}~:\ M(f, \pi) \text{ is feasible}\}.\]
We abuse notation and denote the corresponding quantity for UMF as
simply $f^*$.


\subsection{Inner Approximation to $\Lambda$}

We first show that the maximum UMF $f^*$ is a one-parameter
approximation to the capacity region $\Lambda$. Consider the
following parameter defined in terms of the capacity region as
follows.

\vspace{.1in}

\begin{definition}[$\rho^*$]
For any $\lambda \in \R^{n\times n}_+$, let $ \rho(\lambda)
\stackrel{\triangle}{=} \max_{i} \left\{  \sum_{k=1}^n \lambda_{ik},
\sum_{k=1}^n \lambda_{ki} \right\}.$ Let \\\mbox{$L(x) =
\{\lambda\in\R_+^{n\times n}: \rho(\lambda)\leq x\}$}. Then,  define
$\rho^*$ as
$$\rho^* = \sup\{x\in \R_+: L(x)\subseteq \Lambda\}.$$
\end{definition}
\vspace{.1in} Thus the quantity $\rho^*$ is a parametrization of a
(regular) polyhedral inner approximation to the capacity region
$\Lambda$. It is tight in the sense for any $x>\rho^*$, there is an
infeasible traffic matrix in the set $L(x)$.

Roughly speaking, the following result shows that UMF $f^*$ and
$\rho^*$ are equally good approximations to the capacity region
$\Lambda$.

\vspace{.1in}
\begin{lemma}
\label{lem0} If $U(f)$ is feasible, then any $\lambda\in
\R_{+}^{n\times n}$ such that $\rho(\lambda)\leq nf/2$ is feasible.
\end{lemma}
\begin{proof}
Consider any $\lambda$ such that $\rho(\lambda) \leq nf/2$. Suppose
that $U(f)$ is feasible. Then there exists a transmission scheme such
supports $U(f)$.
We now consider the two stage routing scheme of Valiant and Brenber \cite{valiant}
which routes $U(\rho(\lambda)/n)$ in each stage. Since $U(f)$ is feasible,
any $\lambda$ with $\rho(\lambda) \leq nf/2$ is supportable by time sharing
between the two transmission schemes corresponding to the two stages.
To complete the
proof of the Lemma, we need to describe this two stage routing
scheme.

In the first stage, each node $i$ sends data to all the remaining
nodes uniformly (ignoring its actual destination). Thus, node $i$
sends data to any node $j$ at rate $\sum_{k} \lambda_{ik}/n \leq
\rho(\lambda)/n$.  In the second stage, a node, say $j$, on
receiving data (from the first stage) from any source $i$ sends it
to the appropriate destination. It is easy to see that due to the
uniform spreading of data in the first stage, each node say $j$
routes data at rate $\sum_k \lambda_{ki}/n \leq \rho(\lambda)/n$ to
node $i$ in the second stage. Thus, the traffic matrices routed in
both the stages are dominated by $U(\rho(\lambda)/n)$. That is, the
sum traffic matrix is dominated by $U(2\rho(\lambda)/n)$. Hence, if
$U(f)$ is feasible then $\rho(\lambda)\leq nf/2$ is feasible. This
completes the proof of Lemma \ref{lem1}.
\end{proof}

\vspace{.1in}
\begin{theorem}
\label{lem1}
$f^*$ and $\rho^*$ are related as
\[ \frac{nf^*}{2} \leq \rho^*  \leq nf^* \]
\end{theorem}
\begin{proof} Note that in general, the capacity region $\Lambda$ may not be
closed, and so we need a more careful argument\footnote{We present such a formal
argument only once; similar arguments are implicit in many results that follow}.
We first show that $\frac{nf^*}{2} \leq \rho^*$. By definition of
$\sup$ it follows that for any $\epsilon>0$, $U(f^* - n\epsilon/2)$
is feasible. Hence, from Lemma~\ref{lem0}, any $\lambda\in
\R_{+}^{n\times n}$ such that $\rho(\lambda)\leq nf^*/2 - \epsilon$
is feasible. Hence, again using the definition of $\sup$,
$\rho^*\geq nf^*/2$.

Now for the other bound, assume that $\rho^*  > nf^*$, and $\epsilon
= (\rho^* - nf^*)/2$. Then, by definition of $\sup$ and $\rho$,
$U(nf^* + \epsilon/2)$ is feasible, which is a contradiction. Hence,
it follows that $\rho^*  \leq nf^*$.

\end{proof}

\vspace{.1in}
Thus, bounds on $f^*$ give bounds on $\rho^*$ which differ by at most a factor of 2. Subsequently,
a  scaling law for $f^*$ as a function of $n$ is the same as a scaling law for $\rho^*$,
i.e. $f^* = \Theta(\rho^*)$ as a function of $n$.

The set of all feasible PMF clearly provides an $n$ dimensional
inner approximation to the capacity region, which is, in general,
$n^2$ dimensional. Thus the characterization of the set of feasible
PMFs provides a much better approximation to the capacity region
than that the one-dimensional approximation given by set of feasible
UMF. We next establish the equivalence of UMF with a traffic model
with a randomly chosen permutation flow.

\vspace{.1in}

\subsubsection{UMF and Random Permutation Flow}

In some previous work, (e.g.,~\cite{GK}), the capacity scaling laws
were derived for the case where $n$ distinct source-destination
pairs are chosen at random such that each node is a source
(destination) for exactly one destination (source) and such a
pairing is done uniformly at random over all possible such pairings.
Thus the traffic matrix corresponds to a randomly chosen permutation
flow which is defined as follows.

\vspace{.1in}

\begin{definition}[Permutation Flow]
Let $S_n$ denote the set of permutation matrices in $\mathbb{R}_+^{n\times n}$.
Then the permutation flow corresponding to a permutation $\Sigma\in S_n$ and flow rate
$f \in \R_+$ is given by  $S(f,\Sigma)= f \Sigma$.
\end{definition}
\vspace{.1in}
Many previous works  study the scaling of $\bar{f}$, where $\bar{f}$
is the supremum over the set of $f \in \R_+$  such when a permuation
$\Sigma$ is randomly chosen from $S_n$, the permutation flow $S(f,\Sigma)$
is feasible with probability at least $1-1/n^2$. We now show that
when a permutation flow with flow rate $nf$ and a randomly chosen permutation
is feasible with a high enough probability, then the uniform mulicommodity
flow $U(f)$ can be ``almost" supported when $n$ is large enough.
\vspace{.1in}
\begin{lemma}\label{lem2}
For $\Sigma \in S_n$ chosen uniformly at random, if $(nf)\Sigma$
is feasible with probability at least $1-n^{-1-\alpha}, \alpha > 0$, then
there exists a sequence of feasible rate matrices $\Gamma_n$ such that
\[  \| {U_n(f) - \Gamma_n \|} = O(f n^{-\alpha}) \to 0 \mbox{~as~$n\to\infty$,}\]
where $\| \cdot \|$ denotes the standard 2-norm for matrices\footnote{Given
a matrix $M \in \R^{n \times n}$, the 2-norm of $M$ is $\| M \| = \sup\{ \| Mx \| : x \in \R^n , \|x\| = 1 \}$,
where $\|x\|$ is the $\ell_2$ norm of vector $x \in \R^n$.}, and
$U_n(f)$ is the uniform multicommodity flow for $n$ nodes.
\end{lemma}
\begin{proof}
From the hypothesis of the Lemma, it is clear that for at least
$(1-n^{-1-\alpha})$ fraction of all $n!$ permutations in $S_n$, the
permutation flow $(nf)\Sigma$ is feasible. By definition and
symmetry of permutations, we can write
\[ U_n(f) = \frac{1}{n!}\sum_{i=1}^{n!} (nf)\Sigma_i.\]
Let us define the following indicator function
\[{\bf 1}_i = \left\{\begin{array}{ll}
1&\quad\text{$(nf)\Sigma_i$ is supportable}\\
0&\quad\text{otherwise}
\end{array}\right. \]
Consider a uniform time sharing scheme between all the $n!$
permutation flows. Then the following traffic matrix is supportable.
\[ \Gamma_n = \frac{1}{n!}\sum_{i=1}^{n!} {\bf 1}_i (nf)\Sigma_i\]
Thus
\begin{eqnarray*}
\|{U_n(f) - \Gamma_n\|} &=  &\left|\left|\frac{1}{n!} \sum_{i=1}^{n!}(1- {\bf 1}_i)(nf)\Sigma_i\right |\right|\\
                           &\stackrel{(a)}{\leq} & \frac{1}{n!}\sum_{i=1}^{n!} \left|\left| (1- {\bf 1}_i)(nf)\Sigma_i\right |\right|\\
                           &\stackrel{(b)}{=} & \frac{1}{n!}\sum_{i=1}^{n!}(1- {\bf 1}_i) nf ~
                            \leq ~ \frac{nf}{n!}\frac{n!}{n^{1+\alpha}}\\
                           &= & \frac{f}{n^\alpha}
                            ~\rightarrow 0, ~~\mbox{as $n\to\infty$.}
\end{eqnarray*}
Step (a) uses  triangle inequality for a norm and step (b) uses
$||\Sigma_i||=1$ for any permutation matrix $\Sigma_i$.
\end{proof}

\vspace{.1in} From Lemma~\ref{lem1}, if $U(f)$ is feasible, then
$S(nf/2, \Sigma)$ is feasible for all $\Sigma\in S_n$. Thus, using
an argument identical to that in the proof of Theorem~\ref{lem1}, a
scaling law for $\bar{f}$ is equivalent to a scaling law for $f^*$,
i.e.
$$ f^* = \Theta(n\bar{f}).$$


\subsection{Wireline Networks: PMF Over a Graph}

We briefly review the key results known for PMF on graphs with
fixed edge capacities. These results will be useful in our analysis
for PMF for wireless networks.

Consider a directed graph $G=(V,E)$, where an edge $(i,j)\in E$ has
a capacity $C(i,j)$. Also, for $(i,j)\notin E$, we take $C(i,j)=0$.
 Then for a given $\pi$,
$f_\pi^*$ for graph $G=(V,E)$ is given by the solution of the
following linear program (LP).
\[\begin{aligned}
 \text{maximize}  \qquad & \qquad \qquad f,\\
\text{subject to} \qquad &
\sum_{k: (i,k)\in E}\bbl( x_{ij}(i,k) - x_{ij}(k,i) \bbr)  = f\pi(i)\pi(j), \ 1\leq i,j \leq n,\\
& \sum_{m: (k,m)\in E}\bbl( x_{ij}(k,m) - x_{ij}(m,k) \bbr)  = 0, \ \forall k\neq i,j,\ 1\leq i,j\leq n,\\
& \sum_{i=1}^n\sum_{j=1}^n x_{ij}(k,m)  \leq C(k,m), \ \forall (k,m)\in E,
\end{aligned}\]
where the variables are $f$ and $\{x_{ij}(k,m): (k,m)\in E, i,j,k,m = 1,\hdots, n\}$. The first two
are flow conservation constraints and the third one is the capacity constraint. The total number of variables is
less than $2n^4$ and the total number of constraints is less than $(n^3 + 2n^2)$. Hence, the above LP can
be solved in polynomial time~\cite{karmarkar_1984}.

The well-known max-flow min-cut characterization for a single
commodity flow naturally gives rise to the following question.
Though the maximum PMF $f^*_\pi$ for a given weight
vector can be computed in polynomial time, is there a corresponding
result that relates $f^*_\pi$ and the properties of the graph.
In their seminal paper, Leighton
and Rao \cite{LR88} obtained a characterization of $f^*_\pi$ in
terms of the weighted min-cut of graph. We summarize their
main result below. Let $p_\pi = | \{ i \in V : \pi(i) > 0 \}| $,  denote
the number of nodes for which the corresponding element of $\pi$
is non-zero. Then, without loss of generality we assume that
$\sum_{i=1}^n \pi(i) = p_\pi$.

\vspace{.1in}

\begin{definition}
For the graph $G$ and weight vector $\pi$, define the min-cut by
\[ \Upsilon(G, \pi) = \min_{U\subseteq V} \frac{\sum_{(i,j):i\in U, j\in U^C} C(i,j)}{\pi(U) \pi(U^c)},\]
with notation that $\pi(S) = \sum_{i\in S} \pi(i)$ for any set $S$.
\end{definition}
\begin{theorem}[Theorem 17, \cite{LR99}]
\label{thm:PMF_LR}
In any directed graph $G$, the maximum PMF for weight $\pi$ is related to $\Upsilon(G, \pi)$ as follows:
\[ \Omega\left( \frac{ \Upsilon(G, \pi)} {\log p_\pi}\right ) \leq f^*_\pi \leq  \Upsilon(G, \pi), \]
where the constants for the lower bound do not depend on the graph.
\end{theorem}
\vspace{.1in}
Note that the upper bound follows easily because for a given PMF $f_\pi$, the total flow
from $U$ to $U^C$ is $\pi(U)\pi(U^C)f_\pi$, which has to be less than the total capacity of the
links from $U$ to $U^C$.
The above characterization was crucial to the design of subsequent
approximation algorithms for many NP-hard problems; a summary of these algorithms can be found in~\cite{LR99}. An important case
of the above result is when $\pi(i)=1$ for all $i=1,\hdots, n$, i.e., the special case of uniform mulitcommodity flow.
In this case, we have
\[ \Upsilon(G) = \Upsilon(G, {\bf 1}) = \min_{U\subseteq V} \frac{\sum_{(i,j):i\in U, j\in U^C} C(i,j)}{|U||U^C|},\]
and
\[ \Omega\left( \frac{ \Upsilon(G)} {\log n}\right ) \leq f^* \leq  \Upsilon(G).\]

\vspace{0.1in}

\section{Combinatorial Interference Model}\label{sec2}

A combinatorial interference model defines constraints such that simultaneous data transmissions
over only certain sets of links (or edges) can be successful. This is a simplified abstraction of
a wireless network because in reality whether or not multiple simultaneous data transmissions
are successful depends on the rate of data transmission and the interference power at the various receivers.
We next describe the combinatorial interference model formally and illustrate it with example scenarios where
this abstraction is a reasonable one.

\subsection{Model}
\label{subsec:model}
A combinatorial interference model for a given set of wireless nodes $V=\{1,\hdots,n\}$
defines the following two objects:
\begin{itemize}

\item[(a)] A directed graph $G = (V, E)$ where $E$ is the set of directed links (edges)
over which data can be transmitted.

\item[(b)] For each directed edge $e \in E$, let  $\I(e) = \{ \hat{e} \in E \}$ be the set
of edges (directed links) that interfere with a transmission on link
$e$. Data can be successfully transmitted on link $e$ at rate $W(e)$
if and only if no transmission on any link in $\I(e)$ takes place
simultaneously. In general, the rate $W(e)$ for a given power
constraint can be different for different edges.
The
proof methods and results of the paper will not change
(qualitatively) in this scenario. However, for ease of exposition
we will assume $W(e)=1$\footnote{As long
as $W(e)$ is bounded below and above by a constant, scaling laws
do not change even though the bounds for a given number of nodes $n$ will change.} for all $e \in E$.
\end{itemize}

\vspace{0.1in}
We assume that for every edge $(i,j)\in E$, edge $(j,i)\in E$,  i.e., the graph $G$ is essentially
an undirected graph without the interference constraints given by the sets $\I(e)$'s. This is a reasonable
assumption in many time division and frequency division systems where the channels are reciprocal~\cite{molisch_book}.
The interference sets $\I((i,j))$ and $\I((j,i))$ may not be identical because the transmissions which interfere
with a signal received at node $i$ may not be the same as transmissions which interfere with a signal
received at node $j$.

The above definitions can be used to induce a \emph{dual conflict
graph} as follows.

\vspace{0.1in}
\begin{definition}
The dual conflict graph  is an undirected graph $G^D = (E,E^D)$ with
vertex set $E$ and edge set $E^D$, where an edge $e^D\in E^D$
exists between $e_1$ and $e_2$ if $e_1$ and $e_2$ cannot transmit
simultaneously due to interference constraints.
Thus, each link $e\in E$ is connected to all links in
$\I(e)$.
\end{definition}

\vspace{0.1in}

For the rest of the section, we will suppress the explicit
dependence of all quantities on the combinatorial interference model
parameterized by the graphs $G$ and $G^D$; this helps to simplify
notation. Let us denote the node degree  and the chromatic number\footnote{The chromatic
number of a graph is the minimum number of colors needed to color the vertices of the graph
such that no two nodes  of the graph which are connected by an edge share the same color.}
dual conflict graph $G^D$ by $ \Delta = \max_{e \in E} |\I(e)|$ and
$\kappa$ respectively. Note that $\kappa\leq
(1+\Delta)$. 
Let $\{E_k\}, E_k
\subseteq E$, be the set of all possible link sets that can be active
simultaneously, i.e., simultaneous transmissions
on all the links in $E_k$ at rate $W(e)=1$ are feasible for
the given interference model. Each $E_k$
corresponds to a vector $C_k\in \R^{|E|}$, where $C_k(e) =
{\mathbf 1}_{\{e \in E_k\}} = {\mathbf 1}_{\{e \in E_k\}} $.
Let $\cC$
be the convex-hull of all such vectors $\{C_k\}$. Thus $\cC$ is the set of all
vectors $C$ such that link capacities $C(e)$ (for link $e$) can be obtained
by time-sharing between the $C_k$'s for the given interference model. We then define the capacity
region to be the set of traffic matrices which can be routed over the graph $G = (V,E)$, where each edge
$e$ has capacity $C(e)$ for $C \in \cC$. The formal definition is as follows.

\vspace{0.1in}
\begin{definition}[Capacity Region ($\Lambda$)]
\label{def:cap_region}
The capacity region is the set of traffic matrices $\lambda \in \R^{n\times n}$ such that the following set
of conditions are feasible for some $C\in \cC$:
\begin{equation}
\label{eqn:cap_reg}
\begin{aligned}
&\sum_{k: (i,k)\in E}\bbl( x_{ij}(i,k) - x_{ij}(k,i) \bbr) = \lambda_{ij}, \ 1\leq i,j \leq n,\\
&\sum_{m: (k,m)\in E}\bbl( x_{ij}(k,m) - x_{ij}(m,k) \bbr) = 0, \ \forall k\neq i,j,\ 1\leq i,j \leq n,\\
&\sum_{i=1}^n\sum_{j=1}^n x_{ij}(k,m) \leq C(k,m), \ \forall (k,m)\in E,
\end{aligned}\end{equation}
where $C(k,m) = C(e)$ with $e=(k,m)\in E$; variables are $\{x_{ij}(k,m): (k,m)\in E, 1\leq i,j,k,m \leq n\}$.
\end{definition}
\vspace{0.1in} Thus, the capacity region consists of all traffic
matrices which are feasible using routing and link scheduling
(time-sharing between the sets $\{E_k\}$).  We now illustrate this
capacity region by a couple of special cases corresponding to widely
used models for wireless networks.

\subsubsection{Protocol Model}
The protocol model parameterized by the maximum radius of
transmission, $r$, and the amount of acceptable interference,
$\eta$, is defined in~\cite{GK} as follows.
\begin{itemize}
 \item[(a)] A node $i$ can transmit to any
node $j$ if the distance between $i$ and $j$, $r_{ij}$ is less than
the transmission radius $r$.
\item[(b)] For transmission from node $i$ to $j$ to be successful, no other node
$k$ within distance $(1+\eta) r_{ij}$ ($\eta > 0$ a constant) of
node $j$ should transmit simultaneously.
\end{itemize}
The corresponding definitions of $E$ and $E_D$ follow. A directed
link from node $i$ to node $j$ is in $E$ if $r_{ij}\leq r$. For a
link, $e\in E$, let $e^+$ denote the transmitter and let $e^-$
denote the receiver. Then
\[{\cal I}(e)=\left\{\hat{e}\in E:  r_{\hat{e}^+e^-} \leq (1+\eta)r_{e^+e^-}\right\}.\]
Thus the protocol model is a special case of the combinatorial interference model.
\vspace{.1in}

\subsubsection{SINR Threshold Model}
Assume that all transmissions occur at power $P$, and
the channel gain from the transmitter
of node $j$ to the receiver of node $i$ is given by $h_{ij}$, i.e.,
if node $j$ transmits at power $P$, the received signal power at
node $i$ will be $h_{ij}P$. A signal to interference and noise ratio (SINR) threshold model is parametrized by
a threshold $\alpha$ such that a transmission from node $i$ to node $j$ is successful if and only if the SINR is above
$\alpha$, i.e.,
\[ \frac{P h_{ij}}{ \sum_{k\neq i}Ph_{kj}+ N_0B} \geq \alpha \]
For example, if we assume that each link transmits Gaussian signals and that the Shannon capacity on each link is achievable,
then the threshold is given by $N_0B(2^W-1)$ (assuming $W(e)=W$ for all $e\in E$ as before).

We can define a corresponding combinatorial interference model such that the feasible
simultaneous transmissions defined by the combinatorial interference model are a subset of that
described by the SINR threshold model.
Consider the set of directed links
$E_\gamma$ such that a link, $(i,j)$, from node $i$ to node $j$, is
in $E_\gamma$ if and only if $h_{ji}\geq \gamma$. Also, define
${\cal I}(e) = \{\hat{e}\in E_\gamma: h_{e^-\hat{e}^+ }\geq \beta\}
$.
Then link $e$
can transmit at rate $W$ if no other links in ${\cal I}(e)$ transmit simultaneously,
if and only if $\gamma$ and $\beta$ are such that
\begin{equation}
\label{eqn:interference}   \frac{P h_{e^-e^+}}
{\sum_{\hat{e}\in E_\gamma ,\ \hat{e}\notin {\cal
I}(e)}Ph_{e^-\hat{e}^+} + N_0B} \geq \alpha, \qquad \forall e\in
E_\gamma
\end{equation}
It is easy to see that the above condition is satisfied
if the following condition holds.
\[ \beta \leq \frac{1}{nP}\left(\frac{P\gamma}{2^W-1}-N_0B\right)\]


\subsection{Results}

We now derive results for the combinatorial interference model
which relate the maximum PMF $f_\pi^*$ to spectral properties of the
underlying graphs induced by the interference model. Most of the results
in this section use ideas from known results. While important
in their own right, these results and their proofs motivate the
results for an information theoretic setting for wireless networks
with Gaussian channels. Also, they provide alternate derivations
for known capacity scaling laws in random networks. Towards
the end of this section, we obtain simple bounds on the delay.

\vspace{.2in}

\subsubsection{Bounds on PMF}
For any $C\in \cC$, we denote the maximum PMF
on graph $G$ where each edge $e$ has capacity $C(e)$, by $f_\pi(C)$ , and the corresponding
min-cut by
$$ \Psi_\pi(C) = \min_{S \subset V} \frac{\sum_{(i,j): i \in S, j\in S^c} C(i,j)}{\pi(S)\pi(S^c)}. $$
We denote the corresponding quantities for the special case of UMF by $f(C)$ and $\Psi(C)$, respectively.
Then we have the following lemmas.

\begin{lemma}
\label{lem:cut_continuity}
$\Psi_\pi: \cC \mapsto \R$ is a continuous function for any $\pi \geq 0$.
\end{lemma}
\begin{proof}
Consider a cut $S$ such that $\pi(S)\pi(S^C) >0$. Then, the following is a continuous function of $C$:
$$\frac{\sum_{(i,j): i \in S, j\in S^c} C(i,j)}{\pi(S)\pi(S^c)}.$$
The lemma then follows since the minimum of a finite number of continuous functions is continuous.
\end{proof}

\vspace{0.1in}

\begin{lemma}
\label{lem:pmf_continuity}
$f_\pi: \cC \mapsto \R$ is a continuous function for any $\pi \geq 0$.
\end{lemma}
\begin{proof}
For $\hat{C}\in \cC$ and any $\epsilon>0$ define the set
\[ {\cal B}_\delta = \{C \in \cC : ||C - \hat{C}||< \delta\}.\]
To prove the lemma, we have to show that for any $\epsilon >0$, there exists a
$\delta >0$ such that for all $C\in {\cal B}_\delta$,
$|f_\pi(C) - f_\pi(\hat{C})|<\epsilon$.

For  $\hat{C}\in \cC$ consider
\[ \delta_1 = \min_{(k,m)\in E} \left\{ \alpha \hat{C}(k,m) : \hat{C}(k,m) >0\right \}, \qquad 0<\alpha < 1,\]
and
\[ \delta = \min \left\{ \delta_1, \min_{i,j: \pi(i)\pi(j)>0}\frac{\epsilon}{2\pi(i)\pi(j)}\right \}.\]
Then for any $C \in {\cal B}_\delta $, it follows from~\ref{eqn:cap_reg} that $f_\pi(C) \leq f_\pi(\hat{C}) + \epsilon$.
It only remains to show that $f_\pi(C) \geq  f_\pi(\hat{C}) - \epsilon$.
For this note that $C \succ \underline{C}$  for all $C\in {\cal B}_\delta $, where $\underline{C}$ is as follows:
\[ \underline{C}(k,m) = \left \{
\begin{array}{ll}
0 & \text{C(k,m) = 0}\\
\hat{C}(k,m) - \delta_1 & \text{otherwise}
\end{array}
  \right..\]
Now by scaling all the variables by $(1-\alpha)$ in the LP~(\ref{eqn:cap_reg}) for $\hat{C}$ and
using the monotonicity of $f_\pi(C)$ in $C$, we can see that
$f_\pi(C) \geq (1-\alpha) f_\pi(\hat{C})$ for all $C\in {\cal B}_\delta $. If $f_\pi(\hat{C})=0$,
we are done. If not, choose $\alpha = \min(\epsilon/f_\pi(\hat{C}), 0.5)$, which gives
$f_\pi(C) \geq  f_\pi(\hat{C}) - \epsilon$, and so we are done.

\end{proof}
\vspace{0.1in}

We now define a quantity for the combinatorial interference model
corresponding to the min-cut of a graph.

\vspace{0.1in}
\begin{definition}
The min-cut for the combinatorial interference model is defined as
$$ \Psi_\pi^* = \max_{C \in \cC} \min_{S \subset V} \frac{\sum_{(i,j): i \in S, j\in S^c} C(i,j)}{\pi(S)\pi(S^c)}. $$
\end{definition}

\vspace{0.1in}

Note that $\Psi_\pi^*$ is well defined since $\Psi_\pi(C)$ is a continuous function of $C$, and $\cC$ is closed and
bounded because it is the convex hull of a finite number of points.
The above definition can be interpreted as the min-cut of the graph $G$, where each edge has capacity $C(e)$, and
the vector $C$ is chosen from the set $\cC$ such that it maximizes the min-cut of this graph $G$.
The following result is an extension of Theorem~\ref{thm:PMF_LR} to combinatorial
interference models.
\vspace{0.1in}
\begin{theorem}
\label{thm:UMF_mincut}
$f^*_\pi$ is bounded as
\begin{equation}
\Omega\left(\frac{\Psi_\pi^*}{\log p_\pi} \right) \leq f_\pi^* \leq \Psi_\pi^*.
\end{equation}
\end{theorem}
\vspace{0.1in}
\begin{proof}
Since $\cC$ is closed and bounded, it follows from Lemma~\ref{lem:pmf_continuity} that
there exists $C^*\in \cC$ such that $f^*_\pi = f_\pi(C^*)$. Then, using
Theorem~\ref{thm:PMF_LR}, it follows that
\[
f^*_\pi \leq   \Psi(C^*) \leq \Psi_\pi^*.
\]

Now, from Lemma~\ref{lem:cut_continuity}, there is $\hat{C}\in \cC$ such that
$ \Psi_\pi^* = \Psi_\pi(\hat{C})$. Using Theorem~\ref{thm:PMF_LR}, it follows that
\[f_\pi^* \geq  f_\pi(\hat{C}) = \Omega\left(\frac{\Psi_\pi^*}{\log p_\pi } \right). \]
This completes the proof of Theorem~\ref{thm:UMF_mincut}.

\end{proof}

Note that unlike the case for wireline networks (or equivalently graphs), $f^*$ is a hard
quantity to compute. Also, note that $\Psi$ is a function of both $G$ and the dual graph $G^D$.
We next relate the maximum UMF $f^*$ to spectral properties of graphs $G$ and $G^D$.

\vspace{0.1in}

\begin{definition}
The conductance of a graph $G$ is defined as follows.
\[\Phi(G) = \min_{U\subseteq V, |U|\leq n/2}\frac{\sum_{i\in U, j\in U^C} {\bf 1}_{[(i,j)\in E] }}{|U|},\]
where ${\bf 1}_{[.]}$ is the indicator function.
\end{definition}
\vspace{0.1in}

\begin{corollary}
\label{thm:UMF_cond}
Recall that $\kappa$ is the chromatic number of the dual graph $G^D$.
Then, $f^*$ is related to $\Phi(G)$ as follows.
\[\Omega\left(\frac{\Phi(G)}{\kappa n\log n} \right) \leq f^* \leq \frac{\Phi(G)}{n}.\]
\end{corollary}
\begin{proof}
Consider vertex coloring for the dual graph $G^D = (E,E^D)$.
The chromatic number of $G^D$ is defined to be $\kappa$ and hence we need
$\kappa$ colors for vertex coloring of $G^D$.  Thus we have partitioned
the set $E$ into subsets, say, $E_1,\hdots, E_{\kappa}$ such that the links
in each subset can transmit simultaneously at rate 1.
Now let $C_k(e) = {\bf 1}_{\{e\in E_k\}}$. Then, $C$ corresponding to
uniform time-sharing between the $\kappa$ edge sets $E_1,\dots, E_k$
is given  by
\[ C = \frac{1}{\kappa}(C_1 + \hdots + C_{\kappa}), \]
which is a convex combination of $C_1,\hdots,C_{\kappa}\in \cC$.
Hence, $C(i,j) = 1/\kappa$ for all $i\neq j$, and $C\in \cC$. Then, using Theorem~\ref{thm:PMF_LR} and the definition
of conductance above,
\[ f^* \geq f^*(C) \geq \Omega\left( \frac{\Psi}{\log n} \right) = \Omega\left(\frac{\Phi(G)}{\kappa n\log n} \right).\]
\end{proof}

For the upper bound, note that for any $C\in \cC$, $C \preceq {\bf 1}$, i.e., $C$ is lexicographically
less than ${\bf 1}$, and $f(C_1) \geq f(C_2)$ if $C_1\geq C_2$. Hence, $f^* = \max_{C\in \cC} f(C) \leq f({\bf 1})$. Then, the upper bound follows again
by a straightforward use of Theorem~\ref{thm:PMF_LR} and the definition
of conductance.
\vspace{0.2in}

\subsubsection{Average Delay}

We now provide bounds on the average delay for a class of traffic matrices.
We measure delay in number of hops. We assume that the packet size is small
enough so that the packet delay is essentially equal to the number of hops
taken by the packet. This is similar to the assumptions
in, for example, \cite{GK},\cite{GT}, and \cite{EMPS_journal}.

We restrict ourselves to periodic link scheduling schemes (similar
arguments extend to any ergodic scheduling scheme as well). For
fixed networks, $\cC$ is the convex hull of the set, $\{C_k\}$,
which has a finite cardinality. Hence, any vector in $\cC$ can be
written as a linear combination of the $C_k$'s. Thus to maximize the
supportable uniform multicommodity flow it is sufficient to optimize
over transmission schemes with periodic scheduling of links where
the periodic schedule corresponds to time division between the
$C_k$'s.

We obtain the following general scaling of delay.
\vspace{.05in}
\begin{theorem}\label{thm4}
Let $S(n)$ be the total number of transmissions by the $n$ wireless nodes
on average per unit time\footnote{The quantity $S(n)$ is well defined
since we consider periodic scheduling of links.}. When data is transmitted according to rate matrix
$\lambda\in \Lambda$,  the average delay, $D(n)$, over all packets  scales as
$$ D(n) = \Theta\left(\frac{S(n)}{\bar{\lambda}}\right),
\mbox{~~where $\bar{\lambda} = \sum_{i,j} \lambda_{ij}$.}$$
\end{theorem}
\begin{proof}
Let $\Gamma$ denote the set of all  possible
paths (without cycles) in the network. The amount of flow generated at node $i$
to be transmitted to node $j$ is $\lambda_{ij}$. Let us consider an arbitrary but fixed\footnote{Here,
we consider a fixed deterministic scheme. However, it is easy to see that the result
extends for any randomized scheme as well.}
routing scheme where a fraction $\alpha_{ij}^\gamma$ of
the flow from node $i$ to node $j$ is routed over path $\gamma \in\Gamma$. We assume that
the traffic matrix $\lambda$ is feasible.  Hence, there exists a link scheduling
and routing scheme to support it. The total number of transmissions per unit time at node $l$
is $\sum_{\gamma \ni l}\sum_{i,j} \alpha_{ij}^\gamma\lambda_{ij}$. Hence, the average
number of transmissions per unit time in the entire network, denoted by $S(n)$, is
\[ \begin{aligned}
S(n) \ =\ \sum_{l=1}^n\sum_{\gamma \ni l}\sum_{i,j}
\alpha_{ij}^\gamma\lambda_{ij}
     \ =\  \sum_{\gamma} H^\gamma \sum_{i,j} \alpha_{ij}^\gamma \lambda_{ij}\\
\end{aligned}
\]
where $H^\gamma$ is the number of hops on path $\gamma$.
The total flow over a path $\gamma$ is $\sum_{i,j} \alpha_{ij}^\gamma\lambda_{ij}$, i.e.,
the fraction of total flow over path $\gamma$ is $\sum_{i,j} \alpha_{ij}^\gamma\lambda_{ij}/\bar{\lambda}$.
Hence, the average number of hops traversed by all
packets is given by
\[D(n) = \frac{1}{\bar{\lambda}}\sum_\gamma H^\gamma \sum_{i,j} \alpha_{ij}^\gamma\lambda_{ij} =\frac{S(n)}{\bar{\lambda}}. \]
This completes the proof of Theorem \ref{thm4}.
\end{proof}

\vspace{.2in}

We note that the above result uses very little information about the specific underlying
transmission scheme.
Next, we present an immediate corollary of the above result and Theorem~\ref{thm:UMF_cond} that
bounds the delay for a scheme that maximizes the UMF in the network.

\vspace{.05in}

\begin{cor}
\label{cor2} Since $f^* = \Omega\left(\frac{\Phi(G) }{\kappa n\log n} \right)$, the
corresponding delay scales as $D(n) = O\left( \frac{|E|\log n}{n\Phi(G)} \right)$.
\end{cor}
\begin{proof}
Consider the link scheduling scheme in the proof of
Corollary~\ref{thm:UMF_cond}, where we partition the set of links $E$ into
subsets $E_1,\hdots,E_\kappa$ such that all the links in each subset
$E_i$ can transmit simultaneously. Note that this scheme can support UMF
$f = \Omega\left(\frac{\Phi(G)}{\kappa n\log n}\right)$. For this transmission
scheme, every link transmits
at rate 1 for at most $1/\kappa$ fraction of the time. Hence, we have $ S(n) \leq
\frac{|E|}{\kappa}$. Thus, it follows from Theorem~\ref{thm4} that
$D(n) = O\left( \frac{|E|\log n}{n\Phi(G)} \right)$.

\end{proof}

\vspace{.1in}

\subsection{Computational Methods}

We now describe computational methods to obtain bounds on $f^*$ (the extensions
to PMF are straightforward). As
noted earlier, for wire-line networks, the computation of $f^*$ is
equivalent to solving an LP. However, in a wireless network, the
link capacity is a function of the link schedule. Since, the number
of link schedules is combinatorial, it is a hard problem.
Specifically, the question of checking feasibility of a rate vector
$\lambda$ was proved to be NP-hard by Arikan \cite{Ar84}, that is
there exists an interference model and graph under which checking
feasibility of $\lambda$ is NP-hard.
Motivated by this, here we address the question of providing a
simple computational method to bound $f^*$.

We use ideas of node coloring to induce a link schedule in a way
similar to, for example,~\cite{kodialam_2003}. In particular,
we can obtain an upper bound $f_1^*$ and a lower
bound $f_2^*$ for maximum UMF $f^*$ in polynomial time such that
\[ f_1^* \leq \kappa f_2^* \]
The upper bound can be computed by solving the LP in~(\ref{eqn:cap_reg})
with $C(e)=1$ for all $e\in E$. For the lower bound, since the dual
graph $G^D$ has chromatic number $\kappa$, we can color the nodes of
$G^D$ (which are given by the set $E$ of wireless links) such no
two nodes which share an edge share the same color. This in turn
induces a link scheduling scheme, where each link in $E$ is scheduled
for at least a fraction $1/\kappa$ of time, and the resutling $C$ is
such that $C(e)\geq 1/\kappa$ for all $e\in E$. Again, the lower bound
can be computed by solving the LP in~(\ref{eqn:cap_reg}) with
$C(e)=1/\kappa$ for all $e\in E$. It is easy to see that
$f_1^* \leq \kappa f_2^*$.

Now from Theorem~\ref{thm:UMF_mincut}, we know that
$\Omega\left(\frac{\Psi}{\log n} \right) \leq f^* \leq \Psi$.
Thus, we can now also bound $\Psi$ as
\[ f_2^*     \leq   \Psi \leq O\bl( f_1^* \log n\br ) \]
Thus, the upper and lower bounds differ by at most a factor
of $\kappa \log n$. In addition, using the algorithm
in~\cite{LR88}, we can find a vector $C(e)$ and the
corresponding cut $(U,U^C)$ such that the capacity of
this cut, $\min_{S \subset V} \frac{\sum_{(i,j): i \in S, j\in S^c} C(i,j)}{|S||S^c|}$,
is within a factor $\kappa (\log n)^2$ of $\Psi$.


\subsection{Application}
We now illustrate our results for the combinatorial interference model through an application
to \emph{geometric random graphs}.
The geometric random graph has been widely used to model the topology of wireless networks after the work
of Gupta and Kumar~\cite{GK}. However, it has been a combinatorial object of interest for more than 60 years.
We derive scaling laws for a combinatorial interference model which is more restrictive than the protocol model. Note that
the lower bound hence obtained is also a lower bound for the protocol model. Specifically, the lower bound is weaker by $\log^{2.5}n$
compared to the lower bound obtained in~\cite{GK}. Specifically, we show that the scaling of the lower bound is closely tied
to the known combinatorial properties of geometric random graphs.

We first define the \emph{restricted protocol interference model}.
It is also parameterized by the maximum radius of
transmission, $r$, and the amount of acceptable interference,
$\eta$.
\begin{itemize}
 \item[(a)] A node $i$ can transmit to any
node $j$ if the distance between $i$ and $j$, $r_{ij}$ is less than
the transmission radius $r$.
\item[(b)] For transmission from node $i$ to $j$ to be successful, no other node
$k$ within distance $(1+\eta) r$ (where $\eta > 0$  is a constant) of
node $j$ should transmit simultaneously.
\end{itemize}

\vspace{0.2in}

We now state a version of the well-known Chernoff's bound
for binomial random variables that we use multiple times in this paper.

\begin{lemma}\label{lem:chernoff}
Let $X_1,\dots, X_N$ be i.i.d. binary random variables with $\Pr(X_1 = 1) = p$. Let $S_n = \sum_{k=1}^n X_k$ for
$n=1,\hdots, N$.
Then, for
any $\delta \in (0,1)$
$$ \Pr\left( | S_n - np | \geq \delta n p \right) \leq 2 \exp\left(-\frac{\delta^2np}{2}\right).$$
Specifically, for $\delta = \sqrt{\frac{2L \log n}{np}}$, we have
$$ \Pr\left( |S_n - np| \geq \sqrt{2L np \log n} \right) \leq \frac{2}{n^L}.$$
\end{lemma}

\vspace{.2in}

Consider $n$ wireless nodes  distributed uniformly at random in a unit
square, and the interference model given by the protocol model with transmission
radius $r$. We denote such a wireless network by $G(n,r)$.
It is well-known that for $G(n,r)$ to be connected with high probability,
it is necessary to have $r = \Omega(\sqrt{\log n/n})$.
We take $r = \Theta(\log^{3/4} n/\sqrt{n})$ and prove the following bounds on the maximum UMF, $f^*$,
for the restrictive protocol model;
the lower bound is only $\log^{2.5} n$ weaker than the result of Gupta and Kumar for the protocol model
with $r = \Theta(\sqrt{\log n/n})$.
\begin{lemma}\label{lem:gnrcomb}
For $G(n,r)$, with $r = \Theta(\log^{3/4} n/\sqrt{n})$, maximum UMF is bounded as
$$ \Omega\left(\frac{1}{n^{3/2} \log^{5/2} n}\right) ~\leq~ f^* ~\leq~O\left(\frac{1}{n^{3/2} \log^{3/4} n}\right).$$
\end{lemma}
\begin{proof}
To prove the above bounds, we obtain appropriate upper and lower bounds on the
quantity $\Psi$. These bounds along with  Theorem \ref{thm:UMF_mincut}
imply Lemma \ref{lem:gnrcomb}. To obtain an upper bound on $\Psi$,
we evaluate the cut-capacity for a specific cut-set. For
the lower bound, we first, establish that a grid
graph on $n$ nodes is a sub-graph of $G(n,r)$ and then use the known conductance
of the grid graph.

First, consider the upper bound on $\Psi$. Specifically, consider the square, say $\cS$, of area
$1/9$ (of side $1/3$) that is in the center of the unit square. Let
$S$ be the set of nodes that fall inside this square.
 By definition, we have
 $$ \Psi ~\leq~ \Psi(S) ~=~ \sup_{C\in \cC} \frac{\sum_{i\in S, j\in S^c} C(i,j)}{|S||S^c|}.$$
Therefore, it is sufficient to required obtain bound on $\Psi(S)$.

Corresponding to
node $i$, define a random variable $X_i \in \{0, 1\}$  which is $1$ is $i$
is in $S$, and 0 otherwise.
Since nodes
are placed uniformly and independently at random in the unit area square,
$X_i$ are i.i.d. binary random variable with $\Pr(X_i = 1) = 1/9$. Now,
$\sum_{i=1}^n X_i$ is the number of nodes in $S$. Using Lemma~\ref{lem:chernoff} with
$\delta = 0.009$, it follows that for large enough $n$,
$|S| \in (0.1n,0.2n)$ (and so $|S^c| \in [0.8n, 0.9n]$) with probability at least
$1-n^{-4}$. Now, consider squares $\cS^0, \cS^1$ of sides $1/3 + 2r$
and $1/3 -2r$ respectively with their centers being the same as
that of $\cS$. That is, $\cS^1 \subset \cS \subset \cS^0$. Let
$\cA^0 = \cS^0 - \cS$ and $\cA^1 = \cS - \cS^i$.
Thus, $\cA^1$ is a strip of width $r$ surrounding $\cS$ of
and $\cA^0$ is a strip of width $r$ on the boundary and inside $\cS$.
Since, $r = \Theta(\log^{3/4} n/\sqrt{n})$, it can be easily shown that
$\cA^0$ and $\cA^1$ is $\Theta(r)$.

Now, nodes that are
in $S$ (i.e. physically inside $\cS$) can only be connected
to those nodes in $S^c$ that lie in $\cA^1$. Similarly, nodes
in $S^c$ that are connected to nodes $S$ must lie in $\cA^1$.
Thus, nodes that can communicate across the cut $(S, S^c)$ must
lie within a region of area $\Theta(r)$. For the protocol model,
if a node transmits, nodes within distance $r(1+\eta)$ of the receiver
must not
transmit. That is, each transmission effectively silences nodes
within an area of $\Theta(r^2)$.
Thus, at any given time, the maximum number
of simultaneous transmissions between $S$ and $S^c$ is $\Theta(1/r)$.
This along with $|S|, |S^c| =\Theta(n)$ implies that
$$ \Psi ~\leq~ \Psi(S) ~= ~ \frac{O(1/r)}{\Theta(n^2)} = O\left(\frac{1}{n^2 r}\right)~=~O\left(\frac{1}{n^{3/2} \log^{3/4} n}\right).$$

For the lower bound, we identify
a grid subgraph of $G(n,r)$
with $r = \Theta(\log^{3/4} n/\sqrt{n})$.
Consider a grid graph $G_n$ of $\sqrt{n}\times\sqrt{n}$ nodes
with each node connected to one of its four neighbors (with
suitable modifications at the boundaries). The nodes of $G_n$ are placed in
a uniform manner in a unit square; each node is at a distance
$1/\sqrt{n}$ from its neighbors.
Now consider a minimax matching between
nodes of $G_n$ and $n$ randomly placed nodes in the unit square,
where a minimax matching is a perfect matching between the $n$
nodes of $G_n$ and the nodes of $G(n,r)$ with maximum
length minimized. Leighton and Shor \cite{LS86} established
that the maximum edge length in a minimax matching, say $r^*$, is
$\Theta(\log^{3/4} n/\sqrt{n})$ with probability at least $1-1/n^4$.
Now we identify the subgraph $G^\prime_n$ (with grid graph
structure) of $G(n,r)$ as follows. $G^\prime_n$ has all $n$ nodes.
Consider the minimax matching between $G_n$ and $G(n,r)$. If a node
of $G(n,r)$ is connected to node number $m$ of $G_n$, then renumber
it as $m$ to obtain nodes of $G^\prime_n$. Now by setting
$r \geq r^* + 2/\sqrt{n}$, clearly a node $m$ and $m'$ are connected
in $G^\prime_n$ if they are connected in $G_n$. Thus, we have established
that $G_n \subset G^\prime_n$. Now, we will focus only on the edges
of $G(n,r)$ that belong to $G^\prime_n$ and provide them with positive capacity by an appropriate
communication scheme that is feasible for the restricted protocol model.
For this, note that in $G(n,r)$
each node is connected to at most $O(\log^{3/2}n)$ nodes with probability
at least $1-1/n^4$ (using Chernoff's bound and Union bound) for large enough
$n$. Hence, using a simple TDMA scheme based on vertex coloring of $G(n,r)$,
each node gets to transmit once in every $\Theta(1/\log^{3/2} n)$ time slots.
This transmission can be along any outgoing edge.  Since, we are interested
in providing positive capacity to only at most $4$ outgoing edges,
we have established that there is a simple TDMA scheme which provides
$\Theta(1/\log^{3/2} n)$ capacity to each edge of a grid subgraph of
$G(n,r)$. To complete the proof, we recall that
the conductance of a grid graph is $\Theta(1/\sqrt{n})$ \cite{info_gossip}.
That is,
$$ \Phi(G_n) = \min_{S} \frac{\sum_{i\in S, j\in S^c} {\mathbf 1}_{\{(i,j) \in E\}}}{|S||S^c|} = \Theta\left(\frac{1}{n^{3/2}}\right).$$
Now, putting all the above discussion together we have the following.
\begin{eqnarray}
\Psi & = & \sup_{C\in\cC} \min_{S\subset V} \frac{\sum_{i \in S, j \in S^c} C(i,j)}{|S||S^c|} \nonumber \\
     & \geq & \Phi(G_n) \Theta\left(\frac{1}{\log^{3/2} n}\right) \nonumber \\
     & = & \Omega\left(\frac{1}{n^{3/2} \log^{3/2} n}\right).
\end{eqnarray}
In summary, upper and lower bound on $\Psi$ along with Theorem \ref{thm:UMF_mincut}
implies the Lemma \ref{lem:gnrcomb}.
\end{proof}
\vspace{.1in} Now, we discuss briefly delay. In~\cite{EMPS_journal}, delay
was defined as the average number of hops per packet, and the
packet size was assumed to scale to an arbitrarily small value. For any
communication scheme feasible for the protocol model with maximum transmission radius
$r = \Theta(\log^{3/4} n/\sqrt{n})$, the maximum number of transmissions per
unit time is upper bounded as $O(n/\log^{3/2} n)$. Using this and
Theorem \ref{thm4} we obtain the following result immediately.

\begin{lemma}\label{lem:delay}
The delay $D(n)$ for any scheme achieving
$f^* = \Omega\left(\frac{1}{n^{3/2} \log^{5/2} n}\right)$ is bounded
above as
$$D(n)  =  O\left(\sqrt{n}\log n\right).$$
\end{lemma}

\vspace{.2in}


\section{Gaussian Fading Channel Model}\label{sec3}

In the previous section, we assumed that the wireless
network was defined by two graphs $G$ and $G^D$.
We extended the results of Leighton and Rao
to wireless networks modeled by a combinatorial interference model;
this mainly exploited the fact that all possible transmission schemes could
be described in terms of routing over a set of capacitated graphs, where
the set of edge capacity vectors belonged to the convex hull of a finite
number of vectors. Thus, in this sense, the inherent \emph{discrete}
nature of the model worked to our advantage.

While the combinatorial interference model
can allow for arbitrary scheduling and routing
schemes, it does not model all the degrees of freedom
in a wireless network. Specifically, the results are
not {\em information theoretic}.
In this section, we provide an information
theoretic characterization of PMF in a wireless
network with Gaussian fading channels.
The techniques for the combinatorial model can be easily extended to obtain a feasible scheme and
a lower bound on the maximum PMF $f^*$. However, for information theoretic upper
bounds we have to work harder, especially to obtain a bound that relates to the lower bound
and allows us to quantify the gap.

Our key contribution is in quantifying the suboptimality of the UMF/PMF for
a \emph{simple feasible} scheme and an upper bound on the UMF/PMF
for an arbitrary network topology, in terms of a simple
graph property. The bound is general when channel side information (CSI) is assumed to be
available only at the receiver.
For AWGN channels, we quantify only for UMF, and when the SNR is low enough.   To the best
of our knowledge, this is the first such result which guarantees that a feasible scheme
achieves rates within a certain factor of an outer bound for an arbitrary graph. We also
illustrate these results through applications.
The results hence obtained are interesting in their own right.

Our main approach is as follows. We construct
two directed capacitated graphs $G^U$ and $G^L$ for
the given wireless network. The graph $G^U$ is such that
the capacity (defined appropriately later) of each
cut in $G^U$ upper bounds the corresponding cut-capacity
in the wireless network. The graph $G^L$ is such that
there exists a communication scheme
that simultaneously achieves the capacity of each edge in
$G^L$, and the ratio of capacity of each cut in $G^U$
and $G^L$ is bounded above by a quantifiable term.
This leads to an approximate characterization of
PMF in an arbitrary wireless network with Gaussian
fading channels. Moreover, the feasible scheme that induces the capacities in $G^L$
supports PMF which is within a quantifiable factor of the optimal.


\subsection{Channel Model}

 This is similar to the model in, for example,~\cite{JVK}. We have
$V=\{1,\dots, n\}$ wireless nodes with transceiver capabilities
located arbitrarily in a plane. Node transmissions happen at
discrete times, $t \in \Z_+$. Let $X_i(t)$ be the signal transmitted
by node $i$ at time $t \in \Z_+$. We assume that each node has a
power constraint\footnote{For notational simplicity we assume that
each node has the same power constraint. The general case, where
each node has different maximum average power can be handled using
identical techniques.} such that
\mbox{$\limsup_{N\rightarrow\infty}\frac{1}{N}\sum_{t=1}^N
|X_i^2(t)| \leq P$.}  Then $Y_i(t)$, the signal received by node $i$
at time $t$, is given by
\begin{equation} \label{eqn:gauss_ch}
\textstyle Y_i(t) = \sum_{k \neq i} H_{ik} X_k(t) + Z_i(t),
\end{equation}
where $Z_i(t)$ denotes a complex zero mean white Gaussian noise
process with independent real and imaginary parts with variance 1/2
such that $Z_i(t)$ are i.i.d. across all $i$. Let $r_{ij}$ denote the
distance between nodes $i$ and $j$. Let $H_{ik}(t)$ be such that
$$ H_{ik}(t) = \sqrt{g(r_{ik})} \hat{H}_{ik}(t), $$
where $\hat{H}_{ik}(t)$ is a stationary and ergodic zero mean
complex Gaussian process with independent real and imaginary parts
(with variance 1/2). It models channel fluctuations due to frequency
flat fading. Also, $g(\cdot)$ is a monotonically decreasing function
that models path loss with $g(x) \leq 1$ for all $ x \geq 0$. We
assume also that the $\hat{H}_{ik}(t)$'s are independent.

\subsection{Graph Definitions}

Consider the following two graphs induced by a wireless network of
$n$ nodes:
\begin{itemize}
\item[(1)]~$K_n$ is the fully
connected graph with node set $V$;
\item[(2)]
$G_r$ is the graph where
each node $i \in V$ is connected to all nodes that are within a
distance $r$ of $i$. Let $E_r$ denote the edge set of $G_r$.
Let $\Delta(r)$ be the maximum vertex degree of $G_r$. Finally, define
$$ r^* = \min\{r : ~G_r ~\text{~is connected}\}. $$
\end{itemize}

\vspace{0.1in}

\subsection{Preliminaries}
In the analysis in this section, we utilize the following two simple lemmas.

\begin{lemma}\label{lem:ineq1}
Given $x_i \in (0,1), 1\leq i\leq N$,
$$ \textstyle \sum_{i=1}^N \log (1+\sqrt{x_i}) \leq \sqrt{2N} \sqrt{\sum_{i=1}^N \log(1+x_i)}.$$
\end{lemma}
\begin{proof}
For any $x \in (0,1)$, $x/2 \leq \log (1+x) \leq x$, so
\begin{eqnarray}
\textstyle \sum_{i=1}^N \log (1+\sqrt{x_i}) & \leq & \textstyle \sum_{i=1}^N \sqrt{x_i} \leq  \sqrt{N} \sqrt{\sum_{i=1}^N x_i}  \label{eq:leq2x}\\
& \leq & \textstyle \sqrt{2N} \sqrt{\sum_{i=1}^N \log(1+x_i)},\nonumber
\end{eqnarray}
where (\ref{eq:leq2x}) follows from Cauchy-Schwarz inequality.
\end{proof}

\vspace{0.2in}

\begin{lemma}
\label{lem:ineq2}
For any $x \geq 0$, $\alpha \in (0,1)$,
$\frac{1}{\alpha} \log (1 + \alpha x) \geq \log (1+x)$.
\end{lemma}
\begin{proof}
    Define $f(x) = \frac{1}{\alpha} \log (1+ \alpha x) - \log (1+x)$. Note that $f'(x) \geq 0$ for $x \geq 0$ and $f(0) = 0$.
\end{proof}
\subsection{Results}
We obtain bounds on the maximum PMF for three different cases:
\begin{itemize}
\item[(1)] fading channel with AWGN, and channel side information (CSI) available only
at the receiver,
\item[(2)] deterministic (no fading) additive white Gaussian noise (AWGN) channel,
 and
\item[(3)] fading channel with AWGN, and CSI available at both the transmitter and the receiver.
\end{itemize}
The exact bounds for the above cases are different, but the analysis and bounding techniques are similar.

\vspace{0.3in}

\subsubsection{Random Fading with Rx-only CSI}
We first obtain bounds on the PMF for Gaussian channels with random fading under the assumption
that CSI is available at the receiver, but not the transmitter. We then relate the bounds for PMF, and show
that the gap can be quantified well, and under very general assumptions. We note that this is the case for which we can obtain the
strongest results.
\vspace{0.1in}
\begin{theorem}
\label{thm:coh_bounds}
With channel state information (CSI) only at receivers, $f^*_\pi$ is
bounded as follows:
\[\begin{aligned}
&f^*_\pi \leq  \min_{S\subset V} \frac{\sum_{i\in S, j\in S^c}
\E ( \log(1+ P |H_{ji}|^2 )) }{\pi(S)\pi(S^c)}, \\
&f^*_\pi = \Omega\left(\sup_{r\geq r^*, \ \eta\geq
0}\left[\frac{1}{1+\Delta(r)\Delta(r(1+\eta))}\right]  \times
\left[\min_{S\subset V}\frac{\sum_{i\in S, j\in S^c} {\bf 1}_{(i,j)\in
E_r}\E\log\left( 1 + \frac{P|H_{ji}|^2}{1+nPg(r(1+\eta))}\right ) }{\log
p_\pi \pi(S)\pi(S^c)} \right] \right).
\end{aligned}
\]
\end{theorem}
\vspace{.1in}

Theorem \ref{thm:coh_bounds} provides bounds on $f^*$ which relates to the ``cut capacity" of appropriate capacitated graphs. Specifically, \emph{we can compute the information theoretic upper bound (for any PMF) in polynomial time using flow arguments, and by solving an LP as detailed in Sec.~\ref{subsec:info_comp}}.
However, it is not clear how {\em tight} these bounds are. We now quantify the \emph{gap} between
the upper and lower bounds.

\begin{corollary}
\label{cor:fading}
For any $r\geq r^*$, denote $\delta(r) = \max_{i}\sum_{j: r_{ij} \geq r} Pg(r_{ij})$. Then,
$$
\Omega\left(\frac{\Upsilon }{(1+\Delta^2(r)) (1+\delta(r)) \log p_\pi} \right) \leq f_\pi^* \leq (1+\gamma(r))\Upsilon , $$
where
$$\Upsilon = \min_{S\subset V}\frac{\sum_{i\in S, j\in S^c:  r_{ij} \leq r}
\E \log(1+P |H_{ij}|^2)}{\pi(S)\pi(S^C)}, $$
and
$$ \gamma(r) = \max_{S\subset V: \pi(S), \pi(S^C)>0} \frac{\sum_{i\in S, j\in S^c:  r_{ij} > r} \E \log(1+P |H_{ij}|^2)}
{\sum_{i\in S, j\in S^c:  r_{ij} \leq r} \E \log(1+P |H_{ij}|^2)} .$$
\end{corollary}

\vspace{0.2in}

Note that both $\gamma(r)$ and $\delta(r)$ are decreasing functions of $r$, while $\Delta(r)$ is an increasing function
of $r$. Also, since power typically decays as $1/r^a$ for $2\leq a\leq 6$, while  for uniformly distributed networks
$\Delta(r)$ grows only linearly with $r$, the decay of $\gamma(r)$ and $\delta(r)$ is much faster than the growth of $\Delta(r)$.
Hence, for $r$ large enough the \emph{gap} is dominated by the term $\log p_\pi(1+\Delta(r)^2)$.
Specifically, assume that there exists an $\epsilon > 0$ such that the graph $\hGe =
(V, \hEe)$ is connected, where $\hEe = \{(i,j) : \E \lf[\log \lf(1+
P |H_{ij}|^2\rf)\rf] \geq n^{-\epsilon/2}\}$. Then the above bound for UMF reduces to~\cite{ISIT_paper}
$$
\Omega\left(\frac{\mbox{\sf min-cut}_R}{\Delta^2(r_\epsilon) \log n} \right) = f^* = O(\mbox{\sf min-cut}_R),
$$
where $\mbox{\sf min-cut}_R = \min_{S\subset V}\frac{\sum_{i\in S, j\in S^c:  r_{ij} \leq r_\epsilon}
\E \log(1+P |H_{ij}|^2)}{|S||S^C|}$, and $r_\epsilon$ is such that $\delta(r_\epsilon)\leq \frac{1}{n^{1+\epsilon}}$.

\vspace{0.2in}

\begin{proof}[Theorem \ref{thm:coh_bounds}]
We first prove the upper bound.
Following the steps in the proof of Theorem~2.1 in~\cite{JVK} and using $\left(1+\sum_{i=1}^n \alpha_i\right) \leq \prod_{i=1}^n (1+\alpha_i)$ for $\alpha_i>0$, we obtain that for $\lambda \in \Lambda$,
\begin{eqnarray}
\textstyle \sum_{i \in S, j \in S^c} \lambda_{ij} & \leq & \max_{Q_S\succeq 0, (Q_S)_{ii}\leq P} \E[\log \det (I + H_S Q_S
H_S^*)] \nonumber \\
& \leq & \textstyle \sum_{i\in S, j\in S^c} \E \left( \log(1+ P
|H_{ji}|^2) \right). \label{ezx1}
\end{eqnarray}
Now, for any PMF $M = M(f_\pi, \pi)$,  it must be that $\sum_{i\in S, j\in S^C}M_{ij} = f_\pi \pi(S)\pi(S^C)$.
Hence, for any such PMF $M(f,\pi)\in \Lambda$, the upper bound in the Theorem holds.
\vspace{.03in}

To establish the lower bound, we construct a
transmission scheme for which the PMF is greater
than or equal to that in the lower bound. For $r\geq r^*$, consider
the graph $G_r = (V,E_r)$ on the $n$ nodes defined above. We use
$\Delta(r(1+\eta))$ to denote the maximum vertex degree of the graph
$G_{r(1+\eta)}$. Now, consider the following transmission scheme. A
node $i$ can transmit to a node $j$ only if $r_{ij}\leq r$. Also,
when a node $i$ transmits, no node within a distance $r(1+\eta)$ of
the receiver can transmit. Thus, when a link $(i,j)\in E_r$ is
active, at most $\Delta(r(1+\eta))$ nodes are constrained to remain
silent, i.e., at most $\alpha=\Delta(r(1+\eta))\Delta(r)$ links are
constrained to remain inactive. Hence, the chromatic number of the
dual graph is at most $(1+\Delta(r(1+\eta))\Delta(r)$. In addition,
we assume that the signal transmitted by each node has a Gaussian
distribution.
For any given
link that transmits data at a particular time, we treat all other
simultaneous transmissions in the network as interference. Now
focus on any one link, say link $(1,2)$ between node $1$ and $2$,
without loss of generality. We claim the following.
\begin{lemma}\label{lemma:cx1}
For the above scheme, the following rate on link $(1,2)$ is achievable:
\[ \textstyle \lambda_{12} =\  \alpha^{-1} \E \log \left( 1 + \frac{P |H_{21}|^2}{1 + n Pg(r(1+\eta))} \right).\]
\end{lemma}
We prove Lemma \ref{lemma:cx1} later. First we explain how it implies the proof
of Theorem \ref{thm:coh_bounds}.  A similar analysis holds for
other links that $(1,2)$ in $E_r$. Thus, for graph $G_r$
the following rate are achievable on link $(i,j)\in E_r$:
$$ \alpha^{-1} \E \log \left( 1 + \frac{P |H_{ji}|^2}{1 + n Pg(r(1+\eta))} \right),$$
Now given the capacitated graph $G_r$, we can
use classical wireline network based routing algorithms for obtaining
a product multicommodity flow that is lower bounded by the following
quantity:
$$ f_{LB}(r, \eta) = \Omega\left(\left[\frac{1}{1+\Delta(r)\Delta(r(1+\eta))}\right]  \times
\left[\min_{S\subset V}\frac{\sum_{i\in S, j\in S^c} {\bf 1}_{(i,j)\in
E_r}\E\log\left( 1 + \frac{P|H_{ji}|^2}{1+nPg(r(1+\eta))}\right ) }{\log
p_\pi \pi(S)\pi(S^c)} \right] \right).$$
This implies the following lower bound on $f^*_\pi$:
$$ f^*_\pi \geq \sup_{r \geq r^*, \eta \geq 0} f_{LB}(r, \eta).$$
This is precisely the claimed lower bound in the statement of Theorem \ref{thm:coh_bounds}
and thus completing the proof.
\end{proof}

\begin{proof}[Lemma \ref{lemma:cx1}]
We will use the following result, that follows directly from Theorem~1 in~\cite{kashyap}.
\begin{theorem}
\label{thm:jamming} Consider a complex scalar channel where the
output $Y$ when $X$ is transmitted is given by
\[ \textstyle Y = hX + Z + S,\]
where $Z$ is a complex circularly symmetric Gaussian random variable
with unit variance, and $S$ satisfies $\E[S^*S]\leq \hat{P} $. Also,
$h$ is zero mean and i.i.d over channel uses. If $X$ is a complex
zero mean circularly symmetric Gaussian random variable with
$\E[X^*X]= P $, then $ \textstyle I(X;(Y,h)) \geq  \E \log\left ( 1
+ \frac{P|h|^2}{1 + \hat{P}} \right )$.
\end{theorem}

\vspace{0.1in}

We consider a transmission scheme where the signal transmitted over
each link, when active, is a complex zero mean white circularly
symmetric Gaussian with variance $P$.
 Moreover, we assume that the transmissions on
all links are mutually independent. Let $t_1,t_2,\hdots$ denote
times at which link $(1,2)$ is scheduled. Hence, at any such time
$t\in\{t_1,t_2,\hdots\}$, the received signal at node 2 is given by
\[ \textstyle Y_2(t) = H_{21}(t)X_1(t) + \sum_{k\neq 1,2} H_{2k}(t) X_k(t) + Z_2(t).\]
Using the mutual independence of transmissions and zero mean
property along with the construction of the scheduling scheme,
\[ \textstyle \E \left| \sum_{k\neq 1,2} H_{2k}(t) X_k(t) + Z_2(t)  \right |^2 \leq 1 + nPg(r(1+\eta)).\]
From Theorem~\ref{thm:jamming},
\begin{equation} \label{zza}
\textstyle I( X_1(t); (Y_2(t), H_{21}(t))) \geq \E \log \left ( 1 + \frac{P|H_{21}|^2}{(1 +n Pg(r(1+\eta) )} \right).
\end{equation}
Since the channel is assumed to be i.i.d. over channel uses, a random
coding argument can be used to achieve this rate with a probability
of error that goes to zero as the block length goes to infinity.

Combining this with the time-sharing between different sets of links
described above, since each link gets to transmit at least once in $\alpha$
times slots, or at least $1/\alpha$ fraction of the time, it follows that
\[ \lambda_{12}  \geq \alpha^{-1} \E \log \left( 1 + \frac{P |H_{21}|^2}{1 + n
Pg(r(1+\eta))} \right).\]
\end{proof}

\begin{proof}[Corollary \ref{cor:fading}]
Consider any $S$ such
that $\pi(S), \pi(S^C)>0$. Then,
\begin{eqnarray}
& &  \textstyle \mbox{\sf Cut}(S,S^c) = \sum_{i\in S, j\in S^c} \E \left( \log(1+ P
|H_{ji}|^2) \right). ~~~~~~ \nonumber \\
& & \textstyle  ~\leq (1+\gamma(r)) \sum_{i\in S, j\in S^c: r_{ij} \leq r} \E \left( \log(1+ P
|H_{ji}|^2) \right), ~~~~~\label{eq:cut_bound_rxcsi}
\end{eqnarray}
where the second line follows from the concavity of the $\log$ function, Jensen's inequality,  $\log(1+x)\leq x$
for $x>0$ and definition of $\gamma(r)$.
Thus,
\begin{equation}
\frac{\textstyle \mbox{\sf Cut}(S,S^c)}{\pi(S)\pi(S^C)}\leq \Upsilon (1+\gamma(r)).
\end{equation}
The upper
bound then follows from the upper bound in Theorem~\ref{thm:coh_bounds}.

Next, we consider the transmission scheme that led to the lower bound
in~(\ref{zza}) with $\eta = 0$.
Note that in (\ref{zza}), we used the term $n Pg(r(1+\eta))$ as a bound on the interference power. However, here we consider
 the actual interference $I_{ij} = \sum_{k \in V: r_{jk} \geq r}  Pg(r_{jk})$ for a transmission from $i$ to $j$. Note that
 $I_{ij}\leq \delta(r)$.
Now, by Lemma \ref{lem:ineq2}, we have
\begin{equation} \label{x2aa}
\textstyle \E \lf[\log\lf( 1+
\frac{P|H_{ji}|^2}{1 + I }\rf)\rf] \geq
\frac{1}{1+\delta(r)}\textstyle \E\lf(\log(1+ {P|H_{ji}|^2})
\rf).
\end{equation}
Using (\ref{eq:cut_bound_rxcsi}) and (\ref{x2aa}) along with the
lower bound obtained via time-division scheme that led to (\ref{zza}),
the lower bound in Theorem~\ref{thm:coh_bounds} gives us
\begin{eqnarray}
f^* & = & \Omega\lf( \min_{S\subset V} \frac{\sum_{i\in S, j\in S^c: r_{ij} \leq r} \E \left( \log(1+ P
|H_{ji}|^2) \right)}{\pi(S)\pi(S^C)(1+ \Delta(r)^2) (1+\delta(r))\log p_\pi} \rf) \nonumber \\
& = & \Omega\lf( \frac{\Upsilon}{(1+\Delta(r)^2) (1+\delta(r))\log p_\pi}  \rf). \label{fx0}
\end{eqnarray}
\end{proof}

\vspace{.2in}
\subsubsection{Deterministic AWGN Channels}
We now consider an AWGN channel without fading, i.e., we have $\hat{H}_{kj}=1$ w.p.
$1$, $\forall k,j=1,\hdots,n$.
We first obtain the following set of bounds on maximum PMF using standard arguments.
\begin{theorem}
\label{thm:awgn} The
maximum PMF $f^*_\pi$ is bounded as
follows.
\[
f^*_\pi \leq \min_{S\subset V} \frac{2\sum_{i\in S, j\in S^c} \log(1+
\sqrt{P g(r_{ij}}))}{\pi(S)\pi(S^c)},
\]
\[\begin{aligned} &f^*_\pi = \Omega\left(\sup_{r\geq r^*, \ \eta\geq 0}\left[\frac{1}{1+\Delta(r)\Delta(r(1+\eta))}\right]
\times \left[\min_{S\subset V}\frac{\sum_{i\in S, j\in S^c:  r_{ij} \leq r}
\log\left( 1 + \frac{Pg(r_{ij})
}{1+nPg(r(1+\eta))}\right) }{\log p_\pi \pi(S)\pi(S^c)} \right]\right).
\end{aligned}\]
\end{theorem}

\vspace{0.2in}

Next, we present a Corollary of Theorem \ref{thm:awgn} which characterizes the tightness of the above bound for UMF for
low signal to noise ratio (SNR).

\begin{corollary}\label{cor:lbub}
Define $I(r) = \min\{ I > 0: \sum_{j: r_{ij} \geq r} Pg(r_{ij}) \leq I, ~\mbox{for all $i$}\}$;
$r(\delta) = \min\{ r > 0: I(r) \leq \delta\}$ for $\delta > 0$. Then,
$$
\Omega\left(\frac{n}{(1+\delta)\Delta(r(\delta))(1+\Delta(r(\delta))^2)\log n }  ~{\Upsilon}^2   \right) \leq f^* \leq  2\Upsilon + O\left(\frac{\delta}{n}\right), $$
where
$$\Upsilon = \min_{U\subset V} \frac{\sum_{i\in U, j\in U^c:  r_{ij} \leq r(\delta)} \log(1+ \sqrt{P g(r_{ij}}))}{|U||U^C|}.$$
\end{corollary}

\vspace{.1in}

We now present the proofs of Theorem~\ref{thm:awgn} and Corollary~\ref{cor:lbub}.
The main idea in the proof of Theorem~\ref{thm:awgn}
is to neglect interference to upper bound achievable rates on links, and
to construct a transmission scheme to induce achievable rates on the links. In particular the scheme
that we construct consists of time sharing between multiple transmission schemes, each of which
enables direct transmissions between nodes that are separated by at most distance $r$. Then the lower
bound on $f^*$ is obtained by routing over graph $G_r$, where each edge has a capacity given by this
time division scheme.

\begin{proof}[Theorem \ref{thm:awgn}]
We first prove the upper bound. In order to bound the sum-rate across
each given cut, we refer to the proof of the max-flow min-cut lemma
in \cite{XK}, which yields for any $S \subset V$ and $\lambda \in \Lambda$,
$$
\textstyle \sum_{i \in S, j \in S^c} \lambda_{ij} \leq \sum_{j \in S^c} \log (1 + \E(|\tilde{X}_j|^2)),
$$
where $\tilde{X}_j = \sum_{i \in S} \sqrt{g(r_{ji})} \, X_i$.
 We therefore deduce that
\begin{eqnarray*}
\lefteqn{\textstyle \sum_{i \in S, j \in S^c} \lambda_{ij}}\\
&\leq & \textstyle \sum_{j \in S^c} \log
[1 + \sum_{i,k \in S} \sqrt{g(r_{ji}) \, g(r_{jk})} \,
|\E(X_i \overline{X_k})| ]\\
& \leq & \textstyle \sum_{j \in S^c} \log [ 1 + P (\sum_{i \in S}
\sqrt{g(r_{ji})})^2],
\end{eqnarray*}
since $|\E(X_i \overline{X_k})| \leq \sqrt{P_i P_k} \leq P$. Finally,
we obtain
\[\begin{aligned}
\textstyle \sum_{i \in S, j \in S^c} \lambda_{ij} &\textstyle \leq \sum_{j \in S^c} 2 \log (1 + \sqrt{P} \sum_{i \in S} \sqrt{g(r_{ji})})\\
& \textstyle \leq \sum_{i \in S, j \in S^c} 2 \log (1 + \sqrt{P g(r_{ji})}).
\end{aligned}\]
Now, for any PMF $M = M(f_\pi, \pi)$,  it must be that $\sum_{i\in S, j\in S^C}M_{ij} = f_\pi \pi(S)\pi(S^C)$.
Hence, for any such PMF $M(f,\pi)\in \Lambda$, the upper bound in the Theorem holds.

To establish the lower bound, we construct a
transmission scheme for which the PMF is greater
than or equal to that in the lower bound. For $r\geq r^*$, consider
the graph $G_r = (V,E_r)$ on the $n$ nodes defined above. We use
$\Delta(r(1+\eta))$ to denote the maximum vertex degree of the graph
$G_{r(1+\eta)}$. Now, consider the following transmission scheme. A
node $i$ can transmit to a node $j$ only if $r_{ij}\leq r$. Also,
when a node $i$ transmits, no node within a distance $r(1+\eta)$ of
the receiver can transmit. Thus, when a link $(i,j)\in E_r$ is
active, at most $\Delta(r(1+\eta))$ nodes are constrained to remain
silent, i.e., at most $\Delta(r(1+\eta))\Delta(r)$ links are
constrained to remain inactive. Hence, the chromatic number of the
dual graph is at most $(1+\Delta(r(1+\eta))\Delta(r)$. In addition,
we assume that the signal transmitted by each node has a Gaussian
distribution. Then, subject to the maximum average power constraint,
for any node pair $i, j$, such that $r_{ij} \leq r$,
the following rate is achievable from $i \to j$:
\begin{equation}
\label{cap_lb} \lambda_{ij} \geq  \frac{\log\left( 1 +
\frac{Pg(r_{ij})}{ 1 + nPg(r(1+\eta))}\right )}{1
+ \Delta(r)\Delta(r(1+\eta))}.
\end{equation}
Note that the interference is due to at most $n$ nodes and all the interfering nodes are at least a distance $r(1+\eta)$ away from the receiver. We now consider routing over the graph $G_r$, where each edge $(i,j)$ has capacity $\lambda_{ij}$. The lower bound
then follows from the lower bound in Theorem~\ref{thm:PMF_LR}.
\end{proof}

\vspace{0.2in}

\begin{proof}[Corollary \ref{cor:lbub}]
Consider any cut defined by $(S,S^c)$. Due to the symmetry of the upper bound in Theorem~\ref{thm:awgn}, without loss of generality, assume $|S| \leq n/2$.
Consider any $\delta$ such that $r(\delta)\geq r^*$. Then,
\begin{eqnarray}
& &  \textstyle \mbox{\sf Cut}(S,S^c) =  \sum_{i\in S, j\in S^c} \log (1+\sqrt{Pg(r_{ij})})~~~~~~ \nonumber \\
& & \textstyle  ~=  \sum_{i\in S, j\in S^c :  r_{ij} \leq r(\delta)} \log \lf(1+\sqrt{Pg(r_{ij})}\rf) +
\sum_{i\in S, j\in S^c :  r_{ij} > r(\delta)} \log \lf(1+\sqrt{Pg(r_{ij})}\rf)\nonumber \\
& & \textstyle ~\leq  \sum_{i\in S, j\in S^c :  r_{ij} \leq r(\delta)} \log \lf(1+\sqrt{Pg(r_{ij})}\rf) + |S|\delta,~~~~~\label{eq:leq3}
\end{eqnarray}
where the last step follows from the definition of $r(\delta)$, and $\log(1+\sqrt{x})\leq x$ for
$x\leq 1$.
Hence, the upper bound in the Corollary follows from the upper bound in Theorem~\ref{thm:awgn}.
Since we assume $Pg(r_{ij})\leq 1$ for all $i$ and $j$, from Lemma~\ref{lem:ineq1}, we have
{\small
\begin{eqnarray}
& & \sum_{i\in S, j\in S^c :  r_{ij} \leq r(\delta)} \log \left[ 1+\sqrt{Pg(r_{ij})} \right ]
  ~~\leq  \sqrt{2 \Delta(r(\delta)) |S| \sum_{i\in S, j\in S^c :  r_{ij} \leq r(\delta)} \log \lf(1+Pg(r_{ij})\rf)}. \label{eq:leq4}
\end{eqnarray}}

For the lower bound, consider the choice of $r = r(\delta)$ and $\eta = 0$ for the
scheme described in the proof of Theorem~\ref{thm:awgn}. Then, the interference during data transmission
from $i$ to $j$, $I_{ij} = \sum_{k \in V: r_{jk} \geq r(\delta)}  Pg(r_{jk})\leq \delta$.
Now, Lemma~\ref{lem:ineq2}, implies that
\begin{equation}
\label{eqn:xyz}
\log \left(1 + \frac{Pg(r_{ij})}{1+I_{ij}}\right)
\geq \frac{1}{1+\delta}\log( 1 +  Pg(r_{ij})).
\end{equation}

Using an appropriately modified lower bound in
Theorem \ref{thm:awgn} for the choice of $r = r(\delta)$, $\eta=0$, it follows that
{\small \begin{eqnarray}
f^* & = &  \Omega\left[ \min_{S\subset V} \frac{\sum_{i\in S,
j\in S^c :  r_{ij} \leq r(\delta)} \log (1+Pg(r_{ij}))}
{(1+\delta)(1+\Delta(r(\delta))^2) \log n|S||S^c|}\right] \nonumber \\
& = & \Omega\left[
\frac{\Upsilon^2 n}{(1+\delta)\Delta(r(\delta))(1+\Delta(r(\delta))^2)\log n }\right], \label{x2}
\end{eqnarray}}
where the second step follows from~(\ref{eq:leq4}). The lower bound
in Theorem~\ref{thm:awgn} then implies the lower bound in the Corollary.
This completes the proof.
\end{proof}

\vspace{0.3in}

\subsubsection{Random Fading with CSI at both Tx and Rx}
We now obtain bounds on the PMF for a Gaussian channel with random fading when CSI is available at
both the transmitter and the receiver.
Qualitatively, these bounds are very similar to the case of deterministic
AWGN channels. The main result is as follows.
\vspace{.1in}
\begin{theorem}
\label{thm:coh_bounds1}
With CSI at both transmitters and receivers, $f^*_\pi$ is bounded as follows.
\[
f^*_\pi \leq \min_{S\subset V}\frac{\sum_{i\in S, j\in S^c} 2
\, \E ( \log(1+ \sqrt{P} \, |H_{ji}|)}{\pi(S)\pi(S^c)}.
\]
The lower bound for the receiver only CSI case is a (weak) lower
bound for this case as well.
\end{theorem}
\vspace{.2in}
\begin{proof}
The upper bound follows again from the proof of Theorem 2.1 in \cite{JVK}, from which we deduce that for
any $\lambda \in \Lambda$,
\begin{eqnarray*}
\lefteqn{\textstyle \hspace{-0.2in}\sum_{i \in S, j \in S^c} \lambda_{ij} \leq \E [\max_{Q\succeq 0, Q_{ii}\leq P} \log \det (I + H_S Q_S H_S^*)]}\\
& \leq & \textstyle \sum_{j \in S^c} \E [\max_{Q\succeq 0, Q_{ii}\leq P} \log (1 + h_j Q_S h_j^*)],
\end{eqnarray*}
where $h_j$ is the $j^{th}$ row of $H$. Since $h_j Q_S h_j^*$
is maximum when $(Q_S)_{ik} \equiv P$ for all $i,k \in S$,
we obtain, following the steps of the proof of Theorem~\ref{thm:awgn},
\begin{eqnarray*}
\textstyle \sum_{i \in S, j \in S^c} \lambda_{ij}
& \leq & \textstyle \sum_{j \in S^c} \E \log (1 + P (
\sum_{i \in S} |H_{ij}|)^2 ]\\
& \leq & \textstyle \sum_{j \in S^c} 2 \, \E \log (1 + \sqrt{P}
\sum_{i \in S} |H_{ij}| )]\\
& \leq & \textstyle \sum_{i  \in S, j \in S^c}
2 \, \E (\log (1 + \sqrt{P} \, |H_{ij}|),
\end{eqnarray*}
so the upper bound on $f^*_\pi$ follows from Theorem \ref{thm:PMF_LR}.
\end{proof}

\subsection{Computational Methods}
\label{subsec:info_comp}
We discuss the implications of the bounds
for the case of CSI availability at the receiver only as stated in Corollary \ref{cor:fading}.
Similar implications follow for the case where CSI is available to both transmitters and
receivers as well.

Corollary~\ref{cor:fading} shows that an upper bound
on $f^*_\pi$ can be obtained via the maximum PMF on graph $K_n$,
where each edge $(i,j)$ has a capacity  $\log(1+P |H_{ij}|^2)$, and there is no interference;
specifically, $\log n$ times the PMF thus computed for $K_n$ is an upper bound on $f^*$.
The lower bound is obtained via routing on $G_{r}$ with edge $(i,j)$
having capacity $\frac{\log(1+P |H_{ij}|^2)}{(1+\Delta^2(r))(1+\delta)}$. Hence, the PMF
on $G_{r(\delta)}$ is a lower bound on $f_\pi^*$.
Both the above computations can be done by solving an LP in polynomial time. Moreover, the ratio
of the bounds is quantified in~\ref{cor:fading}. We note that
such an approximation ratio could be obtained easily for the
combinatorial interference model using node coloring arguments.
The arguments  here are more complicated, as detailed
in the proof of  Corollary~\ref{cor:fading}.


\subsection{Application}

We now apply the information
theoretic characterization of PMF in the previous subsection to
obtain a scaling law for average UMF in a geometric random network
with a fading channel, and when CSI is available at the receivers. The scaling law
we obtain is along similar lines to those that exist in the literature.
Similar bounds can be obtained when CSI is available both
at the transmitter and the receiver or when the channels are AWGN channels.

We consider a geometric random graph model
with a constant node density:~$n$ nodes are placed uniformly
at random in a torus of area $n$ (and not unit area). Thus the
distance between two nodes is a random variable taking values
in $(0,\Theta(\sqrt{n}))$. We assume that all nodes have the
same transmission power equal to $1$, i.e. $P_i = 1$ for all
$1\leq i\leq n$. We state the following result characterizing
 $f^*$.
\begin{lemma}
\label{lem:coh_expl}
Consider the Gaussian channel model with random fading and CSI available only
at the receivers. Let $g(r) = (1+r)^{-\alpha}$, $\alpha > 3$ and $P = 1$. Then
for a geometric random graph with constant node density (described above), the average (over
random position of nodes) $f^*$ is bounded as
$$ \Omega\left(\frac{1}{n^{3/2} \log^{1+\alpha} n}\right) \leq \E[f^*] \leq O\left(\frac{1}{n^{3/2}}\right),$$
if $\Pr(|\hat{H}_{ij}|^2 \geq \beta) \geq \gamma$ for some strictily positive constants
$\beta, \gamma$ (independent of $n$) for all $1\leq i, j \leq n$. (Note that the condition is for
the normalized channel gains $\hat{H}_{ij}$s and not the actual gains $H_{ij}$s.)
\end{lemma}
\begin{proof}
We use Theorem \ref{thm:coh_bounds} to evaluate the bounds. First we
obtain an upper bound by evaluating the bound of Theorem \ref{thm:coh_bounds}
for a specific cut $(U, U^c)$. Then, we  evaluate lower bound by
relating it to an appropriate grid-graph as in Lemma \ref{lem:gnrcomb}.

Now, we consider the upper bound. Consider a horizontal line dividing
the square of area $n$ into equal halves. Let $U$ be set of nodes
that lie in bottom half, and so $U^c$ is the set of nodes that lie in the
top half. From Theorem \ref{thm:coh_bounds}, we have
\begin{eqnarray}
f^*|U||U^c| & \leq & \sum_{i \in U, j \in U^c} \E\log (1+P|H_{ij}|^2) \nonumber \\
  & \leq & \sum_{i \in U, j \in U^c} \log (1+P\E[|H_{ij}|^2]) \nonumber \\
  & =  & \sum_{i \in U, j \in U^c} \log (1+Pg(r_{ij})) \nonumber \\
  & \leq & \sum_{i \in U, j\in U^c} Pg(r_{ij}) \nonumber \\
  & = & \sum_{i\neq j} (1+r_{ij})^{-\alpha} \bone_{\{i \in U\}}\bone_{\{j\in U^c\}}, \label{eq:xx1}
\end{eqnarray}
where we have used Jensen's inequality, $\log (1+x) \leq x$ for all $x \geq 0$,
and the hypothesis of Lemma. Since, the nodes are thrown uniformly at random,
the expectation of each term in (\ref{eq:xx1}) for a pair $(i,j)$ is the same.
Using linearity of expectation, we obtain that
\begin{eqnarray}
\E[\sum_{i\neq j}(1+ r_{ij})^{-\alpha} \bone_{\{i \in U\}}\bone_{\{j\in U^c\}}]
& = & n(n-1) \E[(1+r_{12})^{-\alpha}\bone_{\{1 \in U\}}\bone_{\{2\in U^c\}} ] \nonumber \\
& \leq & O\left(n^2 \int_{1}^{\sqrt{n}}  \int_{r}^{\sqrt{n}} s^{-\alpha} \frac{s ds}{n}
\frac{\sqrt{n} dr}{n}\right) \nonumber \\
& = & O\left(\sqrt{n} \int_1^{\sqrt{n}} \int_{r}^{\sqrt{n}} s^{-\alpha+1} ds dr \right) \nonumber \\
& = & O\left(\sqrt{n}\int_1^{\sqrt{n}} r^{-\alpha+2} dr\right) \nonumber \\
& = & O(\sqrt{n}),\label{lx1}
\end{eqnarray}
where we used the fact that for $\alpha > 3$ the last integral is bounded
above by a constant. The above evaluation can be justified as follows. First
note that $\Pr(1 \in U, 2 \in U^c) = 1/4$. Given $\{1\in U, 2 \in U^c\}$,
node $1$ in the bottom rectangle and node $2$ in the top rectangle
are uniformly distributed. Now, consider a thin horizontal strip of
width $dr$ and length $\sqrt{n}$ at distance $r$ below the
horizontal line dividing the square (and inducing $U, U^c$). The node
$1 \in U$ belongs to this strip with probability $2\sqrt{n} dr/n$. Now,
the node $2$ is at distance at least $r$ from node $1$. Consider a ring
of width $ds$, centered at node $1$'s position and of radius $s \geq r$.
The area of this ring is $2\pi s ds$. The probability of node $2$ being
in this ring is bounded above by $4\pi s ds/n$. When the above described
condition is true, the nodes $1$ and $2$ are at distance $s$. Integrating
over the appropriate ranges justifies the final outcome (\ref{lx1}).

Now, it is easy to see that under
any configuration of nodes, $f^* = O(n^2)$ since $g(r) \leq 1$ for any $r \geq 0$,
$P = 1$ and elementary arguments. Let
event $A = \{ |U||U^c| = \Theta(n^2)\}$. Using Chernoff's bound, it is
easy to see that (with appropriate selection of constants in definition of $A$)
for large enough $n$, we have
$$ \Pr(A) = 1-1/n^6. $$
Using this estimate and bound $f^* = O(n^2)$ we obtain that
\begin{eqnarray}
\E[f^*] & = & \E[f^* \bone_A] + \E[f^* \bone_{A^c}] \nonumber \\
        & \leq & \E[f^* \bone_A] + O\left(\frac{1}{n^4}\right) \nonumber \\
        & = & \Theta\left(\frac{\E[f^* |U||U^c|\bone_{A}]}{n^2}\right) + O\left(\frac{1}{n^4}\right) \nonumber \\
        & \leq & \Theta\left(\frac{\E[f^* |U||U^c|]}{n^2}\right) + O\left(\frac{1}{n^4}\right) \nonumber \\
        & = & O\left(\frac{1}{n^{3/2}}\right).
\end{eqnarray}

Next, we prove the lower bound. For this we construct a graph
with achievable link capacities for which the average $f^*$ is
lower bounded as claimed in the Lemma.
Consider $r = \Theta(\log n)$. Then
the corresponding $G_r$, which is the geometric random graph $G(n,r)$, is
connected with high probability (at least $1-1/n^4$ by
appropriate choice of constants in selection of $r$).
For this choice of $r$, using the Chernoff and Union bounds it follows
that with probability at least $1-1/n^4$,
$$ \Delta(2r) = \Theta\left(\log^{2} n\right).$$
Again, we can identify
a grid graph structure as a subgraph structure of
$G_r$ based on the argument used in Lemma \ref{lem:gnrcomb}.
Denote the edges of this grid sub-graph structure as $\hat{E}$.
We note that $\Theta(1)$ edges are incident on each
of the $n$ nodes that belong to $\hat{E}$ (which is a property of
the grid-graph structure). Next, we design a feasible transmission scheme
for which each edge in $\hat{E}$ can support a transmission rate of
$\Omega(\log^{-\alpha} n)$.

Specifically, we consider a TDMA schedule for the
graph $G_r$ similar to that described in the proof of the lower bound
for Theorem \ref{thm:coh_bounds}.
It is easy to
see that $G_{2r}$ can be vertex colored using
$\Theta(\Delta(2r))$ colors.
We use a randomized
scheme to do  TDMA scheduling
as follows: in each time-slot, each node becomes
tentatively active with probability $1/\Delta(2r)$ and
remains inactive otherwise. If a node becomes tentatively active
and none of its neighbors in $G_{2r}$ is tentatively-active, then
it will become active. Else, it becomes inactive. All
active nodes transmit in the time-slot simultaneously.
It is easy to see that each node transmits for
$\Theta(1/\Delta(2r))$ fraction of the time on an average. The
randomization here is used to facilitate the  computation
of a simple bound on the average interference experienced
by a node due to transmissions by  nodes that are
not its neighbor.

Now under the above vertex coloring, each node
gets to transmit once in $\Theta(\Delta(2r))$ time-slots on average
at power $\Theta(P \Delta(2r)) = \Theta(\Delta(2r))$.
We wish to concentrate on transmissions for
edges that belong to $\hat{E}$, which are a subset of
edges of $G_r$. For any such transmission,
say from $u \to v$ with $(u,v) \in \hat{E}$, according
to the above coloring of $G_{2r}$ no other node within distance $r$ of $v$ is transmits
simultaneously. Also, any node
that is at least distance $r$ away from $v$ can be active
with probability at most $1/\Delta(2r)$. Hence, the average power
corresponding to the interference received
by node $v$, say $I_v$, can be bounded above as follows:
\begin{eqnarray}
 I_v & = & O\left(\sum_{j: r_{vj} > r} \frac{P\Delta(2r)\E[|H_{ij}|^2]}{\Delta(2r)}
 \right) = O\left(\sum_{j: r_{vj} > r} g(r_{vj})\right). \label{eq:xx4}
\end{eqnarray}
where we used the fact that each node transmits at
power $P\Delta(2r) = \Delta(2r)$ for $1/\Delta(2r)$ fraction
of the time and $\E[|H_{ij}|^2] = g(r_{ij})$.
By another application of Chernoff's bound and union bound, it
can be shown that the number of nodes in an annulus around node $v$ with unit
width and radius $R$ for $R \in \N, r \leq R \leq \sqrt{n}$, is
$\Theta(R)$ with probability at least $1-1/n^5$. Then it follows that
$$ \sum_{j: r_{vj} > r} g(r_{vj}) = O\left(\sum_{R=\lfloor r \rfloor}^{\sqrt{n}} g(R) R\right) = O\left(r^{-\alpha+2}\right),$$
where we have used fact that $\alpha > 3$. Using this in
(\ref{eq:xx4}), we obtain that with probability at least $1-1/n^{5/2}$ we
have
\begin{eqnarray}
 I_v & = & O\left(r^{-\alpha+2}\right) = O\left(\log^{2-\alpha} n\right).
 \end{eqnarray}
That is, $I_v \to 0$ as $n\to\infty$ for $\alpha > 3$. Thus, by selection
of large enough $n$, $I_v$ can be made as small as possible.
That is, when transmission from $u\to v$ happens, the average noise
received by node $v$ due to other simultaneous transmission is very
small, say less than $\delta$ for some small enough $\delta >0$.

Given this, the arguments used in Lemma \ref{lemma:cx1} imply that
when $u$ transmits to $v$ at power $P\Delta(2r)$ once
in $\Theta(\Delta(2r))$ time-slot, considering other transmissions
as noise, we obtain that the effective rate between $u \to v$ is
lower bounded as
$$ \lambda_{u\to v} = \Omega\left(\Delta(2r)^{-1} \E\left[\log \left( 1+ \frac{P \Delta(2r) g(r_{uv}) |\hat{H}_{uv}|^2}{1+I_v}\right ) \right]\right).$$
Now, $\hat{H}_{uv}$ is independent of everything else and
$$\Pr(|\hat{H}_{uv}|^2 \geq \beta) \geq \gamma,  $$
for some positive constants $\beta, \gamma$ as per our hypothesis.
Therefore, use of Lemma \ref{lem:ineq2} implies
$$ \lambda_{u\to v} \geq \Omega\left(\Delta(2r)^{-1} \log \left (1+ \frac{P \Delta(2r) g(r_{uv})}{1+I_v}\right) \right).$$
Further, $I_v \leq \delta$ for small enough $\delta$. Therefore,
another use of Lemma \ref{lem:ineq2} implies that
 \[\lambda_{u\to v} \geq \Omega\left(\Delta(2r)^{-1} \log \left( 1+ P \Delta(2r) g(r_{uv})\right) \right).\]
Now
$$ P g(r_{uv})\Delta(2r) = \Omega\left( \log^{2-\alpha} n\right),$$
where we have used the fact that $r_{uv} \leq r$ and $g(\cdot)$ is
monotonically decreasing.  For $x \in (0,0.5)$, $\log (1+x) \geq x/2$.
Therefore, for $\alpha > 3$
$$ \lambda_{u\to v} = \Omega\left(\Delta(2r)^{-1} \log^{2-\alpha} n\right).$$
Since $\Delta(2r) = \Theta(\log^2 n)$,  we have established that
the effective capacity of transmissions for each edge under
the above described TDMA scheme is $\Omega(\log^{-\alpha} n)$.
That is, each edge of $\hat{E}$ gets capacity at least
$\Omega(\log^{-\alpha} n)$. Now recall that a grid graph
with unit capacity has $f^*$ lower bounded
as $ \Omega\left(\frac{1}{n^{3/2} \log n}\right)$.
Hence, using this routing of UMF along edges of $\hat{E}$
with capacity $\Omega(\log^{-\alpha} n)$ we obtain that
$$ f^* = \Omega\left(\frac{1}{n^{3/2} \log^{1+\alpha} n}\right).$$
By careful accounting of probability of relevant events above
and Union bound of events will imply that the above stated lower
bound on $f^*$ holds with probability at least $1-1/n^2$. Since
$f^* \geq 0$ with probability $1$, it immediately implies the
desired lower bound of Lemma:
$$ \E[f^*] = \Omega\left(\frac{1}{n^{3/2} \log^{1+\alpha} n}\right).$$
This completes the proof of Lemma \ref{lem:coh_expl}.

\end{proof}

\vspace{.1in}

\vspace{0.3in}


\section*{Acknowledgment}
The authors would like to thank A. Jovicic and S. Tavildar for a helpful discussion on the information
theoretic min-cut bound for networks with Gaussian channels.
We would like to thank Ashish Goel for sharing his proof about evaluating
mixing time of $G(n,r)$ -- it has influenced the proof of Lemma \ref{lem:gnrcomb}.

\bibliographystyle{IEEEtran}
\bibliography{cap}

\end{document}